\documentclass[12pt]{article}
\usepackage{amsmath,amssymb}

\usepackage{graphicx}
\usepackage{epstopdf}
\usepackage{wrapfig}

\usepackage{hyperref}
\usepackage{cite}

\def\beq{\begin{eqnarray}}
\def\eeq{\end{eqnarray}}

\newcommand{\be}{\begin{equation}}
\newcommand{\ee}{\end{equation}}
\newcommand{\bea}{\begin{eqnarray}}
\newcommand{\eea}{\end{eqnarray}}
\newcommand{\bg}{\begin{gather}}
\newcommand{\eg}{\end{gather}}
\newcommand{\bseq}{\begin{subequations}}
\newcommand{\eseq}{\end{subequations}}

\renewcommand{\ln}{\mathop{\rm ln}\nolimits}

\def\be{\begin{eqnarray}}
\def\ee{\end{eqnarray}}
\def\lb{\label}

\usepackage[font=small,labelfont=bf]{caption}

\begin{document}

\title{\textbf{
Extremal black hole mimickers: 
geometry, stability, and long-delay signal trains 
}}
\vspace{1cm}
\author{ \textbf{Sergey N. Solodukhin  }} 

  \date{}
\maketitle
\begin{center}
    \emph{Institut Denis Poisson UMR-CNRS 7013,
  Universit\'e de Tours,}\\
  \emph{Parc de Grandmont, 37200 Tours, France} \\
\end{center}

{\vspace{-11cm}
\begin{flushright}
\end{flushright}
\vspace{11cm}
}



\begin{abstract}
\noindent
 We present a spacetime that serves as a horizonless mimicker of extremal black holes. Unlike previously studied mimickers of black holes with non-degenerate horizons, whose geometries are of the wormhole type and consist of two asymptotically flat regions connected by a very long throat, the mimicker of an extremal black hole possesses a single asymptotically flat region at one end and an infinitely long throat at the other. In the throat region, the spacetime approaches the direct-product geometry $M_{1,1} \times S^2$, where $M_{1,1}$ is two-dimensional Minkowski spacetime and $S^2$ is the two-sphere. We show that the spacetime is  geodesically complete and has everywhere bounded curvature. 
We also investigate scalar-field dynamics and the fate of the Aretakis instability. Although Aretakis-like modes arise in a local analysis of the throat, they do not lead to a physical instability of the mimicker.
A scalar perturbation initially localized in the throat gives rise, at late times, to a sequence of signals with a characteristic separation $t_{\rm delay}\sim 1/\lambda$, where $\lambda\ll 1$ is a deformation parameter. This timescale is parametrically larger than the echo time 
$t_{\rm echo}\sim \ln(1/\lambda)$ characteristic of non-extremal black-hole mimickers. We refer to this sequence as a {\it long-delay signal train}.
 Geometrically, the periodicity of the long-delay signals is associated with periodic geodesics on $S^2$.
We then extend the construction to extremal Kerr.
In this case, the throat region approaches a generalized product geometry involving  $M_{1,1}$ and a two-dimensional surface $\Sigma_2$. The corresponding signals are expected to form a quasi-periodic signal train, reflecting the quasi-periodic geodesic motion on $\Sigma_2$.
We also comment on the generalization to wormhole mimickers of non-extremal black holes and on the implications of these results for unitarity in the context of the AdS/CFT correspondence.


\end{abstract}


\vskip 1 cm
\noindent
\noindent 
\noindent


\pagebreak

\newpage

\section{Introduction}
\setcounter{equation}0

There is accumulating observational evidence for black holes from three independent lines of inquiry \cite{BHreview}: stellar-dynamical measurements, gravitational-wave observations, and Event Horizon Telescope imaging. Yet the possibility remains that the massive compact objects observed through these different channels could instead be horizonless objects that closely mimic the expected observational properties of black holes. Among the various proposed alternatives, wormhole constructions \cite{Damour:2007ap} are particularly interesting, as they provide a framework that combines a degree of universality with the potential to address fundamental black-hole puzzles, such as the unitarity problem \cite{Solodukhin:2005qy}.

Since most of the existing literature focuses on mimicking the properties of non-extremal black holes, such as Schwarzschild \cite{Damour:2007ap} and Kerr \cite{Bueno:2017hyj}, in this paper we extend the wormhole construction to extremal black holes, both static and rotating. 
As is well known, non-extremal and extremal black holes are geometrically quite different. The distinction is particularly sharp in their Euclidean descriptions. In the non-extremal case, the relevant Euclidean geometry has the form of a cigar, with the tip corresponding to the location of the two-dimensional horizon surface. In the extremal case, the tip is replaced by an infinitely long, narrow throat. The two geometries therefore have different topologies.

The horizonless geometry that mimics an extremal black hole is also qualitatively different from that of the non-extremal case. In the latter, the horizon is replaced by a very long throat connecting two asymptotically flat regions. In the extremal case, by contrast, there is only one asymptotically flat region attached to the throat. The throat is described by the spacetime $M_{1,1}\times S^2$, the direct product of two-dimensional Minkowski spacetime and a two-sphere. This spacetime is known in the literature as the Pleba\'nski-Hacyan spacetime. In the present construction, it describes the infinite throat of the horizonless deformation of the extremal black-hole geometry. The resulting mimicker spacetime is geodesically complete and has bounded curvature, which, in the throat region, is independent of the deformation parameter.

Extremal black holes have recently attracted considerable attention following the discovery by Aretakis \cite{Aretakis:2011ha,Aretakis:2011hc} of an instability that is already present for a scalar field governed by the free wave equation. This instability manifests itself as a power-law growth of transverse derivatives of the field along the horizon. For subsequent developments, see \cite{Lucietti:2012xr}-\cite{Katagiri:2021scx}.

The extremal black-hole mimicker constructed in the present paper is horizonless and therefore does not exhibit the Aretakis instability. However, a different type of instability could potentially arise, raising the natural question of whether the instability of the extremal black hole is preserved in some form under the deformation that replaces the horizon with a horizonless geometry. We address this question and indeed find an instability in the spatial derivatives that bears some resemblance to the Aretakis instability. However, the instability we observe has a rather different origin: it appears to be a manifestation of the non-smoothness of the initial data. Thus, we cannot directly trace one instability to the other along the deformation that replaces the extremal black hole with the horizonless mimicker geometry.

Non-extremal black-hole mimickers have a characteristic potentially observable signature. The radial effective potential in these geometries develops a double-peak structure, with the separation between the two peaks scaling as $\ln(1/\lambda)$, see \cite{Damour:2007ap}. As a result, a perturbation entering the throat can generate an additional signal through scattering off the second peak. This is the so-called echo effect \cite{Cardoso:2016rao}, with the characteristic time delay between the direct signal and the echo scaling logarithmically as $\Delta t\sim\ln(1/\lambda)$.

A class of non-symmetric mimickers was introduced in \cite{Solodukhin:2025opw}, in which the two peaks are no longer symmetric. The echo effect is generically present in this more general class of mimickers. The Love numbers of mimickers are also sensitive to the size of the throat region \cite{Solodukhin:2026vtz}. Depending on whether the perturbations probe the throat, the Love numbers can exhibit different behavior, providing another manifestation of the influence of the throat geometry, analogous in a different context to the echo phenomenon.

The double-peak structure is not present for the extremal black-hole mimicker considered here. Only a single peak, already present in the effective potential of the extremal black hole, remains. Consequently, no echo effect is expected in the extremal case. Instead, a perturbation generated in the throat produces a sequence of signals, with a delay time of order
$\Delta t\sim 1/\lambda$.
For small $\lambda$, this timescale is parametrically longer than the echo delay. We call this sequence a \textit{long-delay signal train}.

The origin of this signal train can be traced to singular terms in the retarded Green function associated with caustics in the spacetime $M_{1,1}\times S^2$, as studied in earlier works \cite{Casals:2012px,Aruquipa:2022rwr}. The singularities in question are either delta functions or $\operatorname{PV}$-singular terms, and they lead to characteristic features in the signal that we discuss below. The individual signals propagate through the non-trivial radial potential and can subsequently emerge into the exterior region, where they may, in principle, be detectable.

The observational data, reviewed, for instance, in \cite{Reynolds:2020jwt}, indicate that astrophysical black holes are rotating and that some of them may be very close to extremality. This motivates us to extend our analysis of black-hole mimickers to the extremal Kerr case. In particular, we focus on the throat geometry of the corresponding mimicker and find a family of periodic and quasi-periodic null geodesics, some of which produce caustics similar to those found in the non-rotating case. These results suggest that a sequence of long-delay signals should also arise in the rotating case.

It should be noted that geometries mimicking non-extremal black holes may arise in theories of gravity with generally nonlocal modifications. Such modifications can emerge upon integrating out quantum matter fields. Some evidence supporting this mechanism for black-hole mimickers was found in four dimensions in \cite{Berthiere:2017tms}, and more explicitly in two dimensions in \cite{Potaux:2021yan,Potaux:2022uxa,Potaux:2023fwm}.
For mimickers of extremal black holes, the situation may be less involved: such configurations may arise as solutions of local gravitational equations coupled to suitable matter fields. 
An example of this type is provided by the gravitating 't Hooft–Polyakov monopole\footnote{We thank M. Volkov for bringing this example to our attention.}. As the effective gravitational coupling approaches its critical value from below, the monopole remains a regular, horizonless configuration, while its exterior geometry approaches that of an extremal  black hole. In the critical limit, an infinitely long throat develops, separating the regular monopole interior from the extremal Reissner–Nordström exterior, for details see \cite{Volkov:1998cc}. This example suggests that extremal black-hole mimickers may arise even within local theories of gravity coupled to suitable matter fields. In the present paper, however, we do not investigate the specific mechanisms underlying the formation of such mimickers, but rather focus on their possible observational and physical consequences that may distinguish them from genuine black holes.

The paper is organized as follows. In Section 2, we introduce the geometry of the mimicker. In Section 3, we analyze scalar perturbations in the throat region and address the issue of potential instabilities. In Section 4, we study the retarded Green function in the throat, focusing on its singular terms, which originate from caustics associated with null geodesics propagating on the two-sphere. We then analyze the resulting long-delay signal train. In Section 5, we discuss the mimicker geometry for the extremal Kerr black hole. We conclude in Section 6.

\section{Geometry of the mimicker}
\setcounter{equation}0

\subsection{The metric}

The basic example of an extremal black hole that we consider here is the extremal electrically charged black hole, for which $q=M$, described by the metric,
\be
ds^2=-(\frac{r-M}{r})^2dt^2+(\frac{r-M}{r})^{-2}dr^2+r^2(d\theta^2+\sin^2\theta d\phi^2)\, .
\lb{1}
\ee
The component $g_{tt}$ has a double zero at $r=M$ that indicates that the horizon at $r=M$ is degenerate. In particular the respective Hawking temperature
vanishes.

A horizonless mimicker of this black hole is a spacetime that outside the horizon looks pretty much as described by metric above up to some small corrections.
However, the mimicker metric does not have a true horizon. This is achieved by an everywhere non-vanishing component $g_{tt}$.
A simple recipe suggested in \cite{Damour:2007ap} how to build such spacetime is to replace $-g_{tt}\rightarrow -g_{tt}+ \lambda^2$ where $\lambda$
is a small deformation parameter. Another possibility  suggested in \cite{Bueno:2017hyj} is to have different masses in $g_{tt}$ and $g_{rr}$.
In the Schwarzschild case these two prescriptions are equivalent (after some rescaling of time coordinate $t$).  
In the present context the second prescription seems to be a bit more elegant. That is why we will use it here.
Thus, the suggested form for the mimicker metric is as follow\footnote{Earlier work on mimickers in extremal case is \cite{Lemos:2008cv}.},
 \be
&&ds^2=-\bar{g}(r)^2dt^2+g(r)^{-2}dr^2+r^2(d\theta^2+\sin^2\theta d\phi^2)\, , \ \ \ M_1>M_2 \nonumber \\
&&g(r)=\frac{r-M_1}{r}\, , \ \ \bar{g}(r)=\frac{r-M_2}{r}\, .
\lb{2}
\ee
One can introduce a dimensionless deformation parameter $ \lambda$ as
 \be
 \lambda=\frac{M_1-M_2}{M_1}>0\, .
\lb{3}
\ee
One then has that $M_2=M_1(1-\lambda)$.
For a mimicker the deformation parameter is supposed to be small, $\lambda\ll 1$.
As is clear from the form of the metric (\ref{2}) the mimicker spacetime  is defined for values  $r\geq M_1$ of the radial
coordinate $r$.  So that the value $r=M_2<M_1$, for which $g_{tt}$ vanishes,  does not belong to the spacetime. The minimal value of  ($tt$) component of
the metric, ${\rm min} (-g_{tt})=\lambda^2$. Hereafter we will use $M$ for $M_1$, so that $M_2=M(1-\lambda)$.

\subsection{The curvature}

The scalar curvature for the metric (\ref{2}),
 \be
R=\frac{2 \lambda}{M^2x^3(x - 1 +  \lambda)}\, , \ \ x=r/M
\lb{4}
\ee
is  bounded as function of  $x\geq 1$ and  $0<\lambda<1$. In particular, at $r=M$ the scalar curvature takes value,
 \be
R=2/M^2\, ,
\lb{5}
\ee
that is independent of  $\lambda$.
Quadratic curvature invariants have a similar property,
 \be
R^{\mu\nu} R_{\mu\nu}=\frac{6\lambda^2x^4 - (16\lambda^2 - 8\lambda)x^3 + (20\lambda^2 - 20\lambda + 4)x^2 - (12\lambda^2 - 20\lambda + 8)x + 4(\lambda - 1)^2}{M^4x^8(x - 1 + \lambda)^2}\, , 
\lb{6}
\ee
where $x=r/M$. For $x=1$ ($r=M$) one finds,
 \be
R^{\mu\nu} R_{\mu\nu}=2/M^4\, ,
\lb{7}
\ee
that is independent of value of $\lambda$.

 It should be noted that this is different from the behavior of the curvature for the wormhole mimickers of the Schwarzschild black hole which scales as $1/\lambda^2$ in the throat. Here we see that the curvature in the throat takes ``classical'' value, that characterizes the initial non-deformed geometry, independent of the deformation.

\subsection{Geodesic completeness}

Consider a radial null geodesic, $r=r(\sigma), \ t=t(\sigma)$, where $\sigma$ is affine parameter along the geodesic, in the spacetime described by metric (\ref{2}).
Equation for geodesic then leads to equations,
 \be
\frac{dt}{d\sigma}=\frac{E}{\bar{g}(r)^2}\, , \ \  \frac{dr}{d\sigma}=\pm \, E\, \frac{g(r)}{\bar{g}(r)}\, ,
\lb{8}
\ee
where $E$ is the energy of the photon and we choose sign $(-)$ for a geodesic that goes from a starting position at $r_0$ to decreasing values of the radial coordinate, $r<r_0$.

The change in the affine parameter $\sigma$ along the geodesic is given by the integral,
 \be
\Delta \sigma=\frac{1}{E} \int_r^{r_0}dr \, \frac{\bar{g}(r)}{g(r)}=\frac{1}{E} \left((r_0-r)+M\lambda \ln\frac{r_0-M}{r-M}\right)
\lb{9}
\ee
In the limit, when the radial  coordinate $r$ approaches value $r\rightarrow M$, one finds that the change in the affine parameter is infinite, $\Delta \sigma\rightarrow \infty$.  

The change in the time $t$ measured by a distant observer is given by the integral,
 \be
\Delta t= \int_r^{r_0} \frac{dr} {\bar{g}(r)g(r)}=(r_0-r)+\frac{M}{\lambda}\ln\frac{r_0-M}{r-M}-\frac{M(1-\lambda)^2}{\lambda}\ln\frac{r_0-M_2}{r-M_2}\, .
\lb{10}
\ee
This quantity also diverges as $r\rightarrow M$. Thus, from the perspective of a distant observer, the null geodesic takes an infinite amount of coordinate time to reach $r=M$.

This consideration can be extended to include non-radial geodesics and the time-like geodesics. Indeed in metric (\ref{2}) a generic geodesic characterized 
by conserved energy $E$ and angular momentum $L$ is described by radial equation
 \be
(\frac{dr}{d\sigma})^2=g^2(\frac{E^2}{\bar{g}^2}-\frac{L^2}{r^2}-\kappa)\, ,
\lb{10-1}
\ee
where $\kappa=0$ for null geodesic and $\kappa=1$ for timelike geodesics. In the limit $r\rightarrow M$ this equation reduces to
 \be
(\frac{dr}{d\sigma})^2\simeq(r-M)^2P^2\, ,  \  \  P^2=\frac{1}{M^2}( \frac{E^2}{\lambda^2}-\frac{L^2}{M^2}-\kappa)\, ,
\lb{10-2}
\ee
provided that $P^2>0$.  Integration of this equation then gives rise to
 \be
\Delta \sigma= \frac{1}{P} \int_r^{r_0}  \frac{dr}{r-M}\simeq -\frac{1}{P} \ln(r-M)\, ,
\lb{10-3}
\ee
that diverges when $r$ approaches $M$. 
In the critical case $P=0$, the leading term in (\ref{10-2}) vanishes, and one must expand the right-hand side of (\ref{10-1}) to the next order, which is generally cubic in $(r-M)$. If the coefficient of this term is positive, integration again shows that the corresponding range of $\sigma$ is infinite. 
If the coefficient is negative, the geodesic cannot approach $r=M$. If it vanishes, higher order terms must be considered. 
 Thus, any causal geodesic that can approach $r=M$ does so only after an infinite interval of the affine parameter (or proper time, for timelike geodesics).

The spacetime described by metric (\ref{2}) has two asymptotic ends: one at $r\rightarrow\infty$ and the other at $r\rightarrow M$. At the first end, the spacetime is asymptotically flat, and geodesic completeness is manifest. At the second end, $r=M$, the divergence of the affine parameter for geodesics approaching the boundary demonstrates that the spacetime described by metric (\ref{2}) over the radial range $M<r<\infty$ is geodesically complete.

\subsection{The throat}

For metric (\ref{2})  it is natural to introduce a new radial coordinate, $\rho$, representing the geometrical distance in the radial direction,
 \be
\rho=\int^r \frac{dr}{g(r)}=r-M+M\ln (\frac{r-M}{M})\, .
\lb{11}
\ee
The region where $r\rightarrow M$ in the mimicker spacetime represents the near horizon region in the black hole spacetime.
In this region the new radial coordinate $\rho\rightarrow -\infty$. In this limit one can resolve the relation (\ref{11}) and express
radial coordinate $r$ in terms of $\rho$,
 \be
r(\rho)=M\left(1+e^{\frac{\rho}{M}}-e^{\frac{2\rho}{M}}+\frac{3}{2}e^{\frac{3\rho}{M}}+\dots \right)\, .
\lb{12}
\ee
Respectively, one has that,
 \be
\bar{g}(\rho)=\lambda+(1-\lambda)\left(  e^{\frac{\rho}{M}}-2e^{\frac{2\rho}{M}} +\frac{9}{2}e^{\frac{3\rho}{M}} +\dots \right)    \, .
\lb{13}
\ee
The region of  large negative values of $\rho$ we will call a {\it throat}.
We see that in the throat the correction terms in (\ref{12})- (\ref{13})
are exponentially small and they can be neglected.  Keeping only the leading (constant) terms we arrive at the following
form for the limiting geometry,
 \be
ds^2=-\lambda^2 dt^2+d\rho^2+M^2(d\theta^2+\sin^2\theta d\phi^2)\, .
\lb{14}
\ee
This metric represents the product spacetime $M_{1,1}\times S^2$, where $M_{1,1}$ is two-dimensional Minkowski spacetime and $S^2$ is a two-sphere of radius $M$. It replaces the product spacetime $AdS_2\times S^2$ that describes the near-horizon geometry of the extremal black hole (\ref{1}).

We also note that the time coordinate in the throat, $t_{\mathrm{th}}=\lambda t$, flows much more slowly than the time coordinate $t$ measured by a distant observer when $\lambda\ll1$. This is a generic feature of black-hole mimickers, reflecting the infinite redshift associated with the horizon of a black hole.

Finally, the curvature of metric (\ref{14}) agrees exactly with the limiting values given by Eqs.~(\ref{5}) and (\ref{7}) for the curvature of metric (\ref{2}). Metric (\ref{14}) therefore provides an accurate approximation to the geometry in the throat region.

We also wish to emphasize the distinction  between mimickers of extremal and non-extremal black holes. In the non-extremal case, the mimicker spacetime consists of two asymptotically flat regions connected by a long, although finite, throat. In contrast, as discussed above, the mimicker spacetime of an extremal black hole has no second asymptotically flat region, and its throat is infinitely long.

\subsection{Non-radial geodesics in the throat region} 

Consider a generic non-radial null geodesic. As always the geodesic motion resides to an equatorial plane, $\theta=\pi/2$.
For the other coordinates, $r=r(\sigma), \ t=t(\sigma), \ \phi=\phi(\sigma)$, one has the equations,
 \be
&&\frac{dt}{d\sigma}=\frac{E}{\bar{g}(\rho)^2}\, , \ \     \frac{d\phi}{d\sigma} = \frac{L}{r(\rho)^2} \, , \ \   
 (\frac{d\rho}{d\sigma})^2= \frac{L^2}{\bar{g}(\rho)^2}\, (\frac{1}{b^2}- W(\rho)) \, , \nonumber \\
 &&W(\rho)=\frac{\bar{g}(\rho)^2}{r(\rho)^2}\, , \ \  \frac{1}{b}=\frac{E}{L}\, .
\lb{15}
\ee
In terms of variable $r$, one has for the effective radial potential,
\be
W(r)=\frac{(r-M_2 )^2}{r^4}\, , \ \  M_2=M(1-\lambda)
\lb{16}
\ee
The derivative of the radial potential,
\be
\partial_\rho W(r(\rho))=-\frac{2(r-M_2 )(r-M)(r-2M_2)}{r^6}\, , 
\lb{17}
\ee
vanishes in 3 points. One of them, $r=M_2$ lies outside  the range of the coordinate $r$. The remaining two, $r=M$ and $r=2M_2$, lie
within the physical range, provided that $0< \lambda\leq 1/2$.  
The point $r=M$ corresponds to the asymptotic end of the throat, where the radial derivative of the potential vanishes as
 $r\rightarrow M$. The second stationary point,
$r=2M_2=2M(1-\lambda)$, is the maximum of the radial potential and determines the location of the  unstable  photon sphere.
Its position differs only slightly from that of the extremal black hole,  $r=2M$. The value of the radial potential at the maximum is
$$W(r=2M_2)=\frac{1}{16M_2^2}=\frac{1}{16M^2(1-\lambda)^2}\, .$$

We see that the radial potential $W(\rho)$ has only one maximum, as in the case of the extremal black hole, and  decreases 
in both asymptotic regions. In the region $\rho\rightarrow -\infty$ ($r\rightarrow M$) the potential approaches a constant value,
$W(r=M)={\lambda^2}/{M^2}$. This imposes constraint on possible values of impact parameter $b\leq M/\lambda$ for a geodesic that reaches the throat.

We again note the difference from the case of a non-extremal black-hole mimicker. In that case, the radial potential exhibits two identical maxima separated by a long, nearly flat plateau. 
In the present case, by contrast, the radial potential has only a single maximum, and its overall shape closely resembles that of the extremal black hole, differing only by corrections of order $\lambda)$.

Asymptotically, for large negative values of $\rho$, the radial potential,
\be
W(\rho)=M^{-2}\left({\lambda^2}+ 2\lambda (1-2\lambda)e^{\frac{\rho}{M}}+\dots \right) \, , 
\lb{18}
\ee
takes a constant value with exponentially small corrections.

In the throat, the geodesic equations (\ref{15}) reduce to a set of simple equations,
\be
\frac{dt}{d\sigma}=\frac{E}{\lambda^2}\, , \ \     \frac{d\phi}{d\sigma} = \frac{L}{M^2} \, , \ \   
 \frac{d\rho}{d\sigma}= \pm \frac{1}{\lambda}\sqrt{E^2-\frac{L^2\lambda^2}{M^2}}\, .
\lb{19}
\ee
The solution is given by
\be
t=\frac{E}{\lambda^2}\sigma+t_0\, ,  \ \  \phi=\frac{L}{M^2}\sigma+ \phi_0\, ,  \ \  \rho=\pm \frac{1}{\lambda}\sqrt{E^2-\frac{L^2\lambda^2}{M^2}}\, \sigma+\rho_0\, ,  \ \ \theta=\pi/2
\lb{20}
\ee
First, we note that the throat admits a continuous family of circular null geodesics, each lying along a great circle of $S^2$. These geodesics correspond to the critical impact parameter $b= M/\lambda$.
Due to the translational symmetry $\rho\rightarrow \rho + {\rm const}$, such a circular orbit exists at every value of $\rho$. 
These circular orbits are neutrally stable: radial perturbations neither grow exponentially nor exhibit restoring oscillations. Instead, owing to the translational symmetry of the throat, a small radial displacement simply shifts the geodesic to a neighboring member of the continuous family of circular orbits.

The rotational symmetry of $S^2$  also gives rise to a family of null geodesics that start from the same point on the sphere but in different azimuthal directions. All such geodesics reconverge at the antipodal point, thereby forming a caustic.

A generic geodesic trajectory in the throat, for $b< M/\lambda$, is a superposition of translational motion along the radial direction and periodic motion along a great circle of the two-sphere $S^2$.  

Let us, deep in the throat, consider two points, with radial coordinate $\rho=\rho_0$ and $\rho=\rho_1$ respectively, both located at the same position on the sphere, 
$\theta=\pi/2$ and $\phi=0$. 
These two points are connected by infinitely many null geodesics. A geodesic with winding number $n$ completes $n$ revolutions around a great circle of $S^2$ before reaching the point $\rho=\rho_1$.
The winding number $n$ is determined by the value of impact parameter $b=L/E$. For a geodesic with $n$ windings,
 the corresponding impact parameter is
\be
b_n^{-2}= \frac{\lambda^2}{M^2} \left(1+  \frac{(\rho_1-\rho_0)^2}{(2\pi M n)^2} \right)\, , \qquad n\geq 1\ldots
\lb{21}
\ee
Accordingly, a light signal emitted at $t_0=0$ from $\rho=\rho_0$ reaches the point $\rho=\rho_1$ at the infinite sequence of times $t_n$,
\be
t_n^2=\lambda^{-2}\left[(\rho_1-\rho_0)^2+(2\pi M n)^2\right],\qquad n\geq 0\ldots
\lb{22}
\ee
These arrival times are extremely large, since they scale as $1/\lambda$.

These geodesics go along same grand circle. However, an interesting phenomenon happens when one considers a congruence of geodesics
that go along all possible grand circles originated  from the same point.
A well-known feature of the spacetime $M_{1,1}\times S^2$ is the presence of caustics. A useful way to understand their origin is to consider the congruence of null geodesics emitted from a single event $p\in M_{1,1}\times S^2$. Such geodesics share the same initial point and same constants of motion $E$ and $L$ but differ in the direction of their initial tangent vector on the $S^2$ factor. Since the spacetime is a direct product, their projections onto $M_{1,1}$ are straight lines, while their projections onto $S^2$ are great circles. All great circles emanating from a given point on $S^2$ reconverge at the antipodal point after traversing an angular distance of $\pi$. At the corresponding affine parameter, the null congruence refocuses, producing a caustic. 

As a consequence, sufficiently  separated points in $M_{1,1}\times S^2$ can be connected by an infinite sequence of null geodesics, each corresponding to a different number of windings around the spherical factor. 
Each additional winding involves the passage through another caustic, giving rise to the well-known four-fold singularity structure of the Green function, which will be discussed later in this paper.

\section{Scalar field and stability  in the throat}
\setcounter{equation}0

\subsection{Wave equation and effective radial potential}
Consider a massless scalar field $\Phi(t, r, \theta, \phi)$ satisfying the wave equation
 \be
\Box \Phi=0\, ,
\lb{23}
\ee 
on the background of the mimicker metric (\ref{2}).
As usual, it is convenient to introduce the tortoise coordinate $z$, defined by
 \be
z=\int^r \frac{dr}{\bar{g}(r)g(r)} = r-M+\frac{M}{\lambda}\ln \frac{(r-M)}{M}-\frac{M_2^2}{\lambda M}\ln \frac{(r-M_2)}{M}
  \, , \ \ \ M_2=M(1-\lambda)
\lb{24}
\ee 
For $\lambda\rightarrow 0$ it approaches the tortoise coordinate in the extremal black hole metric. 
The tortoise coordinate $z$ (\ref{24}) changes from $-\infty$ (deep in the throat) to $+\infty$ (asymptotically flat region).
Deep in the throat, where $r \rightarrow M$, the tortoise coordinate $z=M/\lambda \ln (r-M)/M+\dots=\rho/ \lambda +  \dots$, where $\rho$ was introduced earlier in (\ref{12}). In particular, one finds that, 
$$(r-M)=Me^{\lambda z/M},$$ in the throat region.
In the asymptotically flat region, for large $r$, one finds $$z=r+(M+M_2)\ln r +\dots \, .$$
We decompose 
 \be
 \Phi=e^{-i\omega t} \,  Y_{l,m}( \theta, \phi)  \, \frac{\Psi_{\omega, l, m}(r(z))}{r(z)} \, ,
\lb{25}
\ee
where $Y_{l,m}( \theta, \phi)$ are the spherical functions. The radial function $\Psi$ satisfies a Schrödinger type equation,
 \be
\left(-\partial_z^2     +V_{\rm eff}(z)\right)\Psi_{\omega, l, m} (z)=\omega^2 \Psi_{\omega, l, m} \, ,
\lb{26}
\ee
 where the effective radial potential takes the form,
  \be
V_{\rm eff}(z)=\frac{g\bar{g}}{r} \partial_r (g\bar{g})+\frac{l(l+1)}{r^2}\bar{g}^2  \, .
\lb{27}
\ee
The direct calculation gives
  \be
V_{\rm eff}(z)=\frac{(r-M)(r-M_2)}{r^6}\left(M(r-M_2)+M_2(r-M)\right)+ \frac{l(l+1)}{r^4}(r-M_2)^2  \, ,
\lb{28}
\ee
for the radial potential.
Each term in (\ref{28}) is positive for $r \geq M$, and therefore the effective radial potential is everywhere positive.

It vanishes in asymptotically flat region 
and approaches a constant value with exponentially small corrections,
  \be
V_{\rm eff}(z)= \frac{l(l+1)\lambda^2}{M^2} +\frac{\lambda}{M^2}[\lambda+l(l+1)(1-2\lambda)]e^{\lambda z/M} \, ,
\lb{29}
\ee
deep in the throat.

A straightforward, although somewhat tedious, analysis shows that the effective potential $V_{\rm eff}(z)$ has a single extremum, a maximum, at $r=r_{\rm m}$ located close to $r=2M$, where the potential of the extremal black hole also attains its maximum. 
One finds,
  \be
r_{\rm m} = 2M\left(1-\lambda \, \frac{l(l+1)+3/8}{l(l+1)+3/4}- \frac{\lambda^2}{8}\frac{l(l+1)-3/8}{(l(l+1)+3/4)^2}+\dots\right)\, .
\lb{30}
\ee
Clearly, one has that $r_m<2M$.
The value of the potential at the maximum is
  \be
V_{\rm eff}(r=r_{\rm m})=\frac{1}{32M^2}\, \left( (1 + 2l(l+1)) + (4l(l+1) + 1)\lambda   \right)\, .
\lb{31}
\ee
In the undeformed case $(\lambda=0)$, the radial potential (\ref{28}) also has a minimum at $r=M$. For any small but nonzero deformation parameter $\lambda>0$, however, this minimum shifts to the nonphysical region, $r=M(1-\lambda)<M$, and must therefore be discarded. 

In summary, the effective potential (\ref{28}) of the mimicker has the same qualitative shape as that of the extremal black hole. This is a distinctive feature of the present mimicker, in contrast to the non-extremal black hole, whose effective potential exhibits two symmetric maxima separated by a long, nearly flat plateau.

\subsection{Spectrum}

The extremal black hole spacetime is known, following the work of Aretakis, to exhibit a remarkable instability in its near-horizon region. This instability manifests itself through the polynomial growth, as a function of the retarded null coordinate $u$, of the radial derivatives of a field satisfying the free wave equation. A natural question is how this instability is affected in the mimicker spacetime, which can be viewed as a deformation of the extremal black hole. As we have seen, the mimicker spacetime possesses no event horizon, and therefore the Aretakis instability in its original form is clearly absent. However, the near-horizon region is replaced by an infinite throat, making it natural to ask whether the throat retains some kind of ``remnant'' of the instability that was present before the deformation.

The geometry in the throat is described by the direct-product spacetime $M_{1,1}\times S^2$.  In this region, the radial equation (\ref{26}) simplifies to
\be
\left(-\partial_z^2+\mu_l^2\right)\Psi_{\omega,l,m}(z)
=\omega^2\Psi_{\omega,l,m}(z),
\qquad
\mu_l^2=\frac{l(l+1)\lambda^2}{M^2}\, ,
\label{33}
\ee
where $\mu_l$ (for $l>0$) plays the role of an effective mass. On the other side, the throat is connected to an asymptotically flat region. The effective potential is everywhere positive and approaches a positive constant as $z\rightarrow -\infty$, while it vanishes as $z\rightarrow +\infty$. The corresponding Schrödinger operator is therefore non-negative. Moreover, since the potential vanishes at  $z\rightarrow +\infty$, any solution with $\omega^2\ge 0$ is asymptotically oscillatory and cannot be normalizable, even though it may decay exponentially toward $z\rightarrow -\infty$ when $\omega^2\leq \mu_l^2$. Thus, there are no normalizable eigenfunctions, and the spectrum is purely continuous with $\omega^2\ge 0$. In particular, there are no modes with $\omega^2< 0$ that could grow exponentially in time, and the usual instability associated with negative eigenvalues is absent.

Decomposing $\Psi_{\omega, l, m}(z)\sim e^{ikz}$ we have that
 \be
\omega^2=k^2 +\mu_l^2\, .
\lb{34}
\ee
There could be modes with $k^2=-{\rm k}^2<0$, $e^{{\rm k}z}$, that are normalizable on one end ($z\rightarrow -\infty$). For these modes
$0<\omega^2<\mu_l^2$. On the other end  ($z\rightarrow +\infty$) they become oscillatory.
These are evanescent modes in the throat, not unstable modes. The instability would require $\omega^2<0$, and hence $e^{-|\omega |z}$ for $z\rightarrow +\infty$,
which would correspond to a normalizable bound state and is excluded by the positivity of the full Schrödinger operator.

\subsection{Cauchy problem in the throat and some estimates}

In the throat spacetime $M_{1,1}\times S^2$  we can separate the angular variables,  $\Phi=Y_{l,m}(\theta, \phi) \Phi_{l} (t, z)$. The fields $\Phi_{l} (t, z)$
satisfy the two-dimensional massive field equation,
 \be
(-\partial_t^2+\partial_z^2-\mu_l^2)\Phi_{l} (t, z)=0\, , \qquad  \mu_l^2=\frac{l(l+1)\lambda^2}{M^2}
\lb{34A}
\ee
We remind the reader that deep in the throat $\rho=\lambda z$.
The retarded Green function  for the wave equation on spacetime $M_{1,1}\times S^2$  has the form,
 \be
G_R(t-t',  \chi, \gamma)=\frac{1}{4\pi}\sum_{l=0}^{\infty} (2l+1)P_l(\cos \gamma)G_l(t-t', \chi)       \, , \qquad \chi=z-z'
\lb{35}
\ee
where $ \gamma$ is the geodesic distance on two-sphere $S^2$ and $G_l(t-t', z- z')$ is the retarded Green function on $M_{1,1}$ for Eq. (\ref{34A})  
 \be
G_l(t-t', \chi) =\frac{1}{2} \Theta (t-t'-|\chi |)J_0( \mu_l \sqrt{(t-t')^2-\chi^2}    \, . 
\lb{36}
\ee
Let us consider Cauchy problem for Eq.(\ref{34A})  with  initial conditions, 
 \be
 \Phi_{l}(t=0,z)=f_{l}(z)\, , \ \ \  \partial_t\Phi_{l}(t=0,z)=h_{l}(z)\,  .
\lb{37}
\ee
In terms of the retarded Green function solution to the Cauchy problem is presented as follows,
 \be
 \Phi_{l}(t,z)=\int_{-\infty}^{+\infty} dx \left( G_l(t,   z-x) h_l(x)+   \partial_t G_l( t,   z-x)   f_l( x) \right)
\lb{37A}
\ee

With the help of the Green function (\ref{36}) one finds solution for the Cauchy problem
 \be
 &&\Phi_{l}(t,z)=\frac{1}{2}(f_{l}(z-t)+f_{l}(z+t))-\frac{\mu_l t}{2}\int_{z-t}^{z+t} \frac{J_1(\mu_l \sqrt{t^2-(z-x)^2})}{\sqrt{t^2-(z-x)^2}}\,  f_{l}(x) dx \,  \nonumber \\
 && +\frac{1}{2} \int_{z-t}^{z+t}J_0(\mu_l \sqrt{t^2-(z-x)^2})\, h_{l}(x) dx\, .
\lb{38}
\ee
We note that the solution of the Cauchy problem in the massive two-dimensional case receives contributions from every point inside the light cone of $(t, z)$ rather than only from its boundary, as is the case for a massless four-dimensional field. This important difference should be emphasized, since, when combined with the spherical part, the two-dimensional problem yields the solution of the Cauchy problem for a four-dimensional spacetime $M_{1,1}\times S^2$.

Provided that the initial conditions are given by regular functions $f_l(z)$ and $h_l(z)$, the solution (\ref{38}) is regular throughout the entire $(t,z)$ plane. To demonstrate this, we make several estimates.

We consider compact support initial conditions: functions $f_l(z)$ and $h_l(z)$ are bounded and vanish outside region of finite size, $|z|> R$, and
 \be
|f_l(z)| \leq A \, , \ \  |h_l(z)|\leq B  \, . 
\lb{39}
\ee
Using the standard small- and large-argument estimates for Bessel functions 
\be
J_\nu (x)\simeq \frac{1}{\Gamma(\nu+1)}\Big(\frac{x}{2}\Big)^\nu\, , \  \ x\ll 1
\lb{41-0}
\ee
and
\be
J_\nu (x)\simeq \sqrt{\frac{2}{\pi x}}\cos (x-\frac{\nu \pi}{2}-\frac{\pi}{4})\, , \ \ x\gg 1
\lb{41}
\ee
one obtains the following bounds
 \be
|J_0(\mu_l s)|\leq  \frac{C}{(1+\mu_l s)^\frac{1}{2}} \, , \ \   |J_1(\mu_l s)/s|\leq  \frac{C\mu_l}{(1+\mu_l s)^\frac{3}{2}}  \, ,
\lb{40}
\ee
uniformly for  $s=\sqrt{t^2-(z-x)^2}\geq 0$. 

\bigskip

\subsubsection{ Time-like directions: large $t$ and fixed $z$} 

Let us fixe $z$ and consider the limit of large $t$. In this limit starting with some $t$
the values of  $|z-t|$ and $|z+t|$ exceed value of $R$, i.e. these points go outside the domain where functions $h_l$ and $f_l$ are non-vanishing.  Therefore, the first term in (\ref{38}) vanishes and in the two other terms the integration limits can be replaced with $-R$ and $+R$ respectively.

If the coordinate $z$ lies inside the domain where initial conditions are non-vanishing, $|z|\leq R$,
we have that $s=\sqrt{t^2-(z-x)^2}\geq \sqrt{t^2-4R^2}$ provided $t>2R$.
In this case we get an estimate,
 \be
|\Phi_l(t,z)|<\frac{CR}{(1+\mu_l \sqrt{t^2-4R^2})^\frac{1}{2}}(B+A\frac{\mu_l t}{ (1+\mu_l \sqrt{t^2-4R^2})})         \, .
\lb{42}
\ee
If $t>4R$ then one has that $\sqrt{t^2-4R^2}\geq \frac{\sqrt{3}}{2}t$ that helps to simplify the bound (\ref{42}). Indeed in this case
 \be
&&\frac{1}{(1+ \mu_l \sqrt{t^2-4R^2})^\frac{1}{2}}< \frac{1}{(1+\frac{\sqrt{3}}{2} \mu_l t)^\frac{1}{2}}
 \, , \nonumber \\ 
 &&\frac{\mu_l t }{1+\mu_l \sqrt{t^2-4R^2}}< \frac{\mu_l t }{1+\frac{\sqrt{3}}{2} \mu_l t}<\frac{2}{\sqrt{3}} \, .
\lb{43}
\ee
Thus, the bound (\ref{42}) takes a simpler form,
 \be
|\Phi_l(t,z)|<\frac{CR}{(1+\frac{\sqrt{3}}{2} \mu_l t)^\frac{1}{2}}(B+A\frac{2}{ \sqrt{3})})         \, , \ \ t>4R, \ |z|\leq R
\lb{44}
\ee
or even simpler if we drop $1$ in the square root, we find a bound
 \be
|\Phi(t,z)|<\frac{C'R(A+B)}{\sqrt{\mu_l t}}        \, , \ \
\lb{45}
\ee
 valid for $t>4R$, where $C'=C 2^{3/2}3^{-3/4}$.

Consider now the case when $z$ lies outside the domain where the initial conditions are defined, $|z|>R$. Suppose, for simplicity, that $z>R$ then 
$s=\sqrt{t^2-(z-x)^2}> \sqrt{t^2-(z+R)^2}$ provided $t>z +R$. In this case we get an estimate,
 \be
|\Phi_l(t,z)|<\frac{CR}{(1+\mu_l \sqrt{t^2-(z+R)^2})^\frac{1}{2}}(B+A\frac{\mu_l t}{ (1+\mu_l \sqrt{t^2-(z+R)^2})})         \, ,
\lb{46}
\ee
that is similar to (\ref{42}) but now it depends on $z$. It is valid for any $t>z +R$.
If we demand that
$t> 2(z+R)$ then we can go along same lines as before (\ref{43}), (\ref{44}) and get a bound identical to (\ref{45}) that is thus valid also for $|z|>R$.  This consideration can be extended to $z<-R$. The bound  (\ref{45}) is valid for times $t>2(|z|+R)$.

Combining these two regimes we conclude 
that the bound (\ref{45}) is valid for any fixed $z$ for sufficiently large $t>T(z, R)={\rm max} \{4R, 2(|z |+R)\}$.  
It does not establish, however,  a bound uniform in all $z$, because $T(z, R)$ grows with $|z |$.
Thus, for any fixed $z$, the field $\Phi_l$  decays  at least as $1/\sqrt{t}$ at late times.
The decay power in this bound is independent of the angular momentum $l$, although the coefficient of the bound depends on $\mu_l$
 (and hence generally on $l$).

It might be of interest to analyze large $t$ behavior of (\ref{38}) for  a concrete choice of functions $h_l(x)$ and $f_l(x)$.
Since the $f$-dependent part of (\ref{38}) is $t$-derivative of the $h$-part (with function $h$ replaced with  $f$), it is
sufficient to focus on the $h$-part of the solution (\ref{38}). The simplest functions in the class of compactly supported functions are
\[
h^{(k)}(x)=
\begin{cases}
(x^2-R^2)^k, & |x|\leq R,\\
0, & |x|>R,
\end{cases}
\qquad k=0,1,2,\ldots
\]
For a fixed \(k\), this function has continuous derivatives of orders
$
p=0,\ldots,k-1
$
at the boundary points \(x=\pm R\). In other words,
$h^{(k)}\in C^{k-1}(\mathbb{R})$.

This follows from the asymptotic behavior of the Bessel function (\ref{41}) that  for large $t$ the integral,
 \be
 \int_{-R}^{R}J_0(\mu_l \sqrt{t^2-(z-x)^2})\, h^{(k)}(x)\sim \frac{1}{\sqrt{t}}    \, ,  \ \ t\gg R
\lb{47}
\ee
independently of values of $k$.  This indicates that one can not get a faster time decay  than $\frac{1}{\sqrt{t}} $ in this class of functions.

\bigskip

\subsubsection{ Null directions}  The decay of the solution along the null (light-like) directions is different from that of along the time-like directions. In order to see this consider a light ray given by equation $z(t)=t+z_0$, where $z_0$ is the point on the Cauchy surface $t=0$. This point can be either inside,  $-R\leq z_0\leq R$, or outside  $z_0\leq -R$ of the interval $[-R,R]$ (for $z_0>R$ there is no intersection of the light-cone with the interval $[-R,R]$ and the resulting solution vanishes). The conclusion we make  for the asymptotic behavior of the solution
(\ref{38}) along the ray is the same for both cases although details of the analysis are slightly different. For simplicity we consider the case when $-R\leq z_0\leq R$ so that the light ray starts inside the interval. 
Let us focus on the part in the solution that is due to the initial function $h(x)$. For large $t$ the lower and upper limits in the  integral over $x$ are $z_0$ and $R$ respectively. Notice that $(x-z_0)\geq 0$.
The spacetime interval $s^2=t^2-(z(t)-x)^2=2(x-z_0)t-(x-z_0)^2\simeq 2(x-z_0)t$ for sufficiently large $t$.
We see that in this case the interval $s\sim \sqrt{t}$. This is different from the time-like direction in which $s\sim  t$.
This differences in particular leads to a different behavior of the solution for large $t$.

Let us  focus on the $h$-dependent part, $Q_h(t)=\Phi(t) |_{f=0}$, of the solution (\ref{38}) and take $h^{(0)}(x)=1$ as initial condition function.
Then the integral
 \be
 Q^{(0)}_h(t)=\int^{R}_{z_0}J_0(\mu_l \sqrt{2(x-z_0)t})dx
 =\frac{\sqrt{2(R-z_0)}}{\mu \sqrt{t}}J_1(\mu \sqrt{2(R-z_0)t})   \, .
\lb{48}
\ee
Note that that the integral comes from the end-point $x=R$ of the interval $[-R,R]$ where the initial data are fixed. If we considered $z_0<-R$ then the entire interval
$-R\leq x \leq R$ should be taken in the integral and the result would come from the both ends of the interval.

Using the asymptotic form of the Bessel function (\ref{41}), we find the asymptotic form of the solution,
  \be
 Q^{(0)}_h(t)
 =-\frac{2(R-z_0)^{1/4}}{\mu^{3/2} t^{3/4}}\cos (\mu \sqrt{2(R-z_0)t}+\pi/4)   \, .
\lb{49}
\ee
We see that for this choice of the initial function the solution $Q(t)\sim 1/t^{3/4}$ decays  faster than the $1/t^{1/2}$ behavior found in the case of the time-like directions. 

A similar, although slightly more involved, analysis can be carried out  for initial function $h^{(1)}(x)=x^2-R^2$,
  \be
 &&Q^{(1)}_h(t)=\int^{R}_{z_0}(x^2-R^2)J_0(\mu_l \sqrt{2(x-z_0)t})dx \nonumber \\
 &&=\frac{ 2^{1/4}(R-z_0)^{3/4}}{\sqrt{\pi} \mu^{5/2} t^{5/4}}\sin (\mu \sqrt{2(R-z_0)t}+\pi/4) 
\lb{50}
\ee
for large $t$. We see that, in this case, the solution decays even faster, as $1/t^{5/4}$. 

If the same  pattern continues,  then for  the initial function $h^{(k)}(x)=(x^2-R^2)^k$ the solution $Q^{(k)}_h(t)$ is expected to decay as  $1/t^{(3+2k)/4}$. This suggests that, for an initial function whose derivatives of all orders vanish at the boundary of the interval  $[-R,R]$, i.e. $h(x)\in C^{\infty}_c(\mathbb{R})$, the solution  in the null directions decays faster than any power of $1/t$ for large $t$.  This appears to be consistent with the mathematical study in \cite{Sussman}, see also an earlier work \cite{Winicour:1988aq},  where the statement was extended to  derivatives of the scalar field.

Similar observations can be made for the $f$-part of the solution (\ref{38}), $P_f(t)=\Phi(t) |_{h=0}$. 
Skipping the technical details, we present the results. 

For the initial function $f(x)=h^{(0)}(x)=1$ one finds along the null direction
$z(t)=t+z_0$ (again, for simplicity, we choose $-R<z_0<R$), that for large values of $t$,
  \be
 P^{(0)}_f(t)=\frac{2^{1/4}}{2\sqrt{\pi}(R-z_0)^{1/4}\mu^{1/2}\, t^{1/4}}\, \sin (\mu \sqrt{2(R-z_0)t}+\pi/4)\, . 
\lb{51}
\ee
Thus, the solution decays as $1/t^{1/4}$ in this case. This is slower than the $1/t^{1/2}$ behavior expected from the
large $t$ analyses above at fixed $z$. 

For the initial function $f(x)=h^{(1)}(x)=x^2-R^2$, one finds, for large $t$,
  \be
 P^{(1)}_f(t)=\frac{2^{3/4}(R-z_0)^{1/4}R}{\mu^{1/2}\, t^{3/4}}\, \cos (\mu \sqrt{2(R-z_0)t}+\pi/4)\, . 
\lb{52}
\ee
The solution thus decays as $1/t^{3/4}$. 
More generally, for the initial function  $f(x)=h^{(k)}(x)=(x^2-R^2)^k$ 
the  solution  $ P^{(k)}_f(t)$ is   expected to decay as
$1/t^{(1+2k)/4}$. For an initial function $f(x))\in C^{\infty}_c(\mathbb{R})$  
the corresponding solution in the null directions is therefore 
 expected to decay faster than any power of $1/t$ for large $t$.
We, however, should notice that in the light-like direction the decay of the $f$-part of the solution is slower than that of the $h$-part.
 
 \bigskip

\subsubsection{ Spatial derivative instability} 

We have seen above that the decay of the scalar field along the null directions depends on the degree of regularity of the 
 initial condition functions at the boarder of the interval $[-R,R]$. Despite this decaying behavior of the field itself
 the spatial derivatives of sufficiently high order will diverge. In order to see this let us consider the asymptotic form of the field (\ref{49}). It is a combination of the decaying term $1/t^{3/4}$ and a rapidly oscillating function of $\chi (t, z_0)=\mu \sqrt{2(R-z_0)t}+\pi/4$.  
 Here $z_0$ parametrizes the null ray $z=t+z_0$. Varying $z_0$ therefore amounts to varying the spatial position of the null ray. The derivative of the oscillatory phase is
 \be
 \partial_{z_0} \chi (t, z_0)=-\frac{\mu\sqrt{t} }{\sqrt{2(R-z_0)}}\sim \sqrt{t}\, .
 \lb{D1}
\ee
 Thus, every additional derivative with respect to $z_0$ can increase the power of $t$ by $1/2$.
 Differentiating (\ref{49}) w.r.t. $z_0$ one increases the power of $t$ so that the derivative of sufficiently high order becomes divergent when time $t$ goes to infinity,
  \be
 Q^{(0)}(t, z_0)\sim t^{-3/4}\, , \ \ \partial_{z_0} Q^{(0)}(t, z_0)\sim t^{-1/4}\, , \ \ \partial^2_{z_0} Q^{(0)}(t, z_0)\sim t^{1/4}\, ,
 \lb{D2}
\ee
where we omit bounded oscillatory factors and $z_0$-dependent prefactors.

 A similar behavior one finds for $Q^{(1)}(t, z_0)$,
   \be
 Q^{(1)}(t, z_0)\sim t^{-5/4}\, , \ \ \partial_{z_0} Q^{(1)}(t, z_0)\sim t^{-3/4}\, , 
 \ \ \partial^2_{z_0} Q^{(0)}(t, z_0)\sim t^{-1/4}\, , \ \ \partial^3_{z_0} Q^{(1)}(t, z_0)\sim t^{1/4} .
 \lb{D3}
\ee
Thus, although the scalar field itself decays along the null ray, spatial derivatives of sufficiently high order can grow with time.

This behavior bears a striking resemblance to the Aretakis instability, in which sufficiently high transverse derivatives of a scalar field grow along an extremal horizon. The underlying mechanisms, however, are quite different. In the present case, the growth originates from the endpoint behavior of the initial data and from the $\sqrt{t}$ dependence of the oscillatory phase in the null asymptotics.

More precisely, $Q^{(k)}(t,z_0)$ is the solution corresponding to the initial profile $h^{(k)}(z)$, whose derivatives of order $n<k$ vanish at the boundary of $[-R,R]$, whereas its $k$-th derivative is non-vanishing there. Upon extending $h^{(k)}$ by zero outside the interval, its $k$-th derivative has a jump at $z=\pm R$; consequently, the next distributional derivative contains delta contributions supported at the endpoints. The examples above suggest that this finite-order endpoint regularity controls the order of the first growing spatial derivative.

 Assuming that the asymptotic pattern  $ Q^{(k)}(t, z_0)\sim t^{-(3+2k)/4}$ 
 holds in general, differentiation with respect to $z_0$ gives
\be
\partial^n_{z_0} Q^{(k)}(t, z_0)\sim t^{-(3+2k-2n)/4}\, .
  \lb{D4}
\ee
The exponent becomes positive for $n\geq k+2$. 
 
 This observation indicates that the leading null asymptotic is determined by the first non-vanishing derivative of the initial data at the endpoint. In particular, the null decay depends sensitively on the finite-order regularity of the initial data at the boundary of their compact support.

For smooth compactly supported initial data, all derivatives vanish at the boundary, and the finite-order endpoint mechanism described above is absent. In this case, the solution of the massive Klein--Gordon equation restricted to a null line, together with all of its spatial derivatives, is rapidly decreasing, in agreement with the results of \cite{Sussman}.

Interestingly, the derivative-growth mechanism described above occurs for every fixed $\mu>0$: the mass affects the oscillatory phase and the coefficients, but not the powers of $t$. This is distinct from the Aretakis mechanism, where the relevant conserved quantities for the massive scalar field arise for the special relation $m^2=l(l+1)$ between the mass and the angular momentum eigenvalue. 

As the mimicker spacetime considered here is a deformation of the extremal black hole, it is natural to ask whether the Aretakis instability is preserved, in some form, under this deformation. The analysis presented here shows that, although we find a derivative instability that bears some resemblance to the Aretakis instability, the two phenomena have fundamentally different origins. In the present case, the growth of sufficiently high spatial derivatives is associated with the null asymptotics and the finite-order regularity of the initial data at the boundary of their support. It is not associated with a conserved quantity on a degenerate horizon, nor does it require the special relation between the mass and angular momentum that underlies the Aretakis mechanism. Thus, although the derivative growth found here resembles that of the Aretakis instability, it appears to represent a distinct phenomenon rather than a direct continuation of the Aretakis mechanism in the deformed geometry.

\bigskip

\subsection {   A closer look: the throat versus the extremal horizon}  
To examine more closely what happens to the Aretakis instability in the mimicker,
 let us consider a simple example of two-dimensional Anti-de Sitter spacetime and repeat the Aretakis 
analysis. We consider a scalar field with mass $m=\sqrt{l(l+1)}$ arising from the spherical harmonics on the two-sphere.
The Aretakis type instability in the case of $AdS_2$  was studied in \cite{Lucietti:2012xr}, \cite{Katagiri:2021scx}.

For the mimicker metric we take the following metric,
\be
ds^2=-(r^2+\lambda^2)dt^2+\frac{dr^2}{r^2}\, .
 \lb{D4}
\ee
A small $\lambda$ removes the degenerate horizon at $r=0$ and replaces it with a throat.
The corresponding in-going Eddington-Finkelstein  coordinate is  defined as
\be
v=t+z\, , \ \   z=\frac{1}{\lambda}\ln \frac{r}{\lambda^2+\lambda\sqrt{r^2+\lambda^2}}\, .
 \lb{D5}
\ee
In the coordinates $(v,r)$ the metric becomes
\be
ds^2=-(r^2+\lambda^2)dv^2+2\frac{\sqrt{r^2+\lambda^2}}{r}drdv\, .
 \lb{D6}
\ee
The massive field equation then takes the form
\be
\frac{2r}{\sqrt{r^2+\lambda^2}}\partial_v\partial_r \Phi+\frac{r(2r^2+\lambda^2)}{r^2+\lambda^2}\partial_r\Phi+r^2\partial_r^2\Phi=l(l+1)\Phi\, .
 \lb{D7}
\ee
Following the Aretakis analysis, we seek solutions which are regular at $r=0$ by expanding the field in powers of $r$. To illustrate the relevant behavior, we consider the simple ansatz,
\be
\Phi(v,r)=r^\alpha F(v)\, .
 \lb{D8}
\ee
In the standard Aretakis analysis $\alpha$ is a positive integer. Near $r=0$ the equation (\ref{D7})  reduces to 
\be
\partial_v F(v)=\frac{\lambda}{2\alpha}[l(l+1)-\alpha^2]F(v)\, 
 \lb{D9}
\ee
with the solution
\be
F(v)=F_0e^{\Gamma(\alpha)v}\, , \ \   \Gamma(\alpha)=\frac{\lambda}{2\alpha}(l(l+1)-\alpha^2)\, .
 \lb{D10}
\ee
At first sight, this result appears to imply an exponential instability for $0<\alpha<\sqrt{l(l+1)}$ which would be parametrically stronger than the polynomial growth of transverse derivatives characteristic of the Aretakis instability at an extremal horizon.

However,  this conclusion requires some care. The spacetime (\ref{24}) near $r=0$ approaches Minkowski metric with coordinates $(T,X)$, 
\be
T=\lambda t\, ,  \ \ X=\ln(r/\lambda)\, .
 \lb{D11}
\ee
In terms of these coordinates the solution (\ref{D8}), (\ref{D10}) takes the plane-wave form
 \be
 \Phi(T,X)= e^{-i\Omega T}e^{ikX}\, , \  \  \, 
 \lb{D12}
\ee
where
 \be
k=-\frac{i}{2\alpha}(l(l+1)+\alpha^2)\, , \ \ \Omega=\frac{i}{2\alpha}((l(l+1)-\alpha^2)\, .
 \lb{D12-2}
\ee
These quantities satisfy 
\be
\Omega^2=k^2+l(l+1),
\lb{D13}
\ee
as expected for a massive scalar in the asymptotically Minkowskian throat.

As is clear from this form, for real positive $\alpha$ both $k$ and $\Omega$ are imaginary. 
In particular, for
$0<\alpha<\sqrt{l(l+1)}$, one has $\operatorname{Im}\Omega>0$, corresponding to the apparent exponential growth found above.
These modes decay as $X\to-\infty$, i.e. toward the asymptotically Minkowskian end.
If one formally continued the flat-space approximation toward $X\to+\infty$, the same solution would grow exponentially. However, this continuation is not physical: the Minkowski approximation breaks down in this direction, and the mode encounters the non-trivial radial potential of the full geometry.

Indeed, in terms of coordinates $(t,z)$ the massive field equation 
\be
(-\partial_t^2+\partial_z^2)\Phi=m^2(r^2+\lambda^2)\Phi\, , \ \ m^2=l(l+1)
\lb{D14}
\ee
after substitution $\Phi=e^{-i\omega t}\psi (z)$ becomes a Schr\"odinger type equation
\be
-\partial_z^2\psi+V(z)\psi=\omega^2\psi\, , \ \  V(z)=l(l+1)(r^2+\lambda^2)\, .
\lb{D15}
\ee
The effective potential is everywhere positive and approaches
$l(l+1)\lambda^2$ as $r\to0$, while it diverges as $r\to\infty$.
At the Minkowskian end, physical solutions must obey the appropriate finite-energy condition, while at the asymptotic $AdS_2$ boundary the two independent behaviors are $\psi\sim r^l$ or $\psi\sim r^{-(l+1)}$. 
The standard $AdS_2$ boundary condition selects the normalizable branch, $\psi\sim r^{-(l+1)}$. 
Since the potential in  (\ref{D15}) is positive  no physical mode with $\omega^2<0$, and therefore no exponentially growing mode, can satisfy the global boundary conditions.

Thus, the apparent exponentially growing solutions obtained from the local Aretakis-like analysis are not global unstable modes of the mimicker. They arise from treating the near-throat region locally and do not by themselves determine the global spectrum. The positive-definite radial operator, together with the appropriate boundary conditions at the Minkowskian end and at the asymptotic $AdS_2$ boundary, excludes an exponentially growing physical mode. This highlights the essential difference between the mimicker throat and an extremal horizon: the local Aretakis-like behavior survives, but it does not lead to a genuine Aretakis instability.

 \section{Caustics and  the long-delay  signal train }
\setcounter{equation}0

\subsection {Sequence of singular terms in retarded Green function}
In the previous section we considered the propagation of modes at fixed value of the angular momentum $l$.
Let us know combine this with the summation over $l$ and discuss the propagation of the entire scalar field signal in the throat
as function of the spacetime coordinates $(t, z, \theta, \phi)$ on the spacetime $M_{1,1}\times S^2$.
The corresponding retarded Green function $G_R(t-t', z-z', \gamma)$ is given by equation (\ref{35}).
Here $\gamma$ is the geodesic distance on two-sphere $S^2$ between two points with coordinates $(\theta, \phi)$ and $(\theta', \phi')$,
  \be
\cos \gamma= \cos \theta  \cos \theta'+  \sin\theta   \sin\theta' \cos( \phi-\phi')\, , \ \ 0\leq \gamma \leq \pi
\lb{53}
\ee
With equation (\ref{36}) we find for four-dimensional retarded Green function (\ref{35}),
  \be
G_R(t-t', z-z', \gamma)=\frac{1}{4\pi}\Theta(t-t'-|z-z'|)\sum_{l=0}^{\infty} (2l+1)P_l(\cos\gamma)J_0(\frac{\lambda}{M}\sqrt{l(l+1)}s)
\, ,
\lb{54}
\ee
where $s=\sqrt{(t-t')^2-(z-z')^2}$.
Our goal now is to find the singular part of the Green function. For that we first expand both Legendre function and the Bessel function for large values of $l\gg 1$,
  \be
P_l(\cos(\gamma)) \simeq 
\sqrt{\frac{\gamma}{ \sin\gamma}} J_0((l+\frac{1}{2})\gamma)
\, ,
\lb{55}
\ee
and the large argument expansion for the Bessel function is given in (\ref{41}). Noting that 
$\sqrt{l(l+1)}=\sqrt{(l+1/2)^2-1/4}\simeq l+1/2$ we find for the product of the Legendre function and the Bessel function appearing in 
(\ref{54}),
   \be
&&\frac{1}{4\pi}\sum_{l=0}^{\infty} (2l+1)P_l(\cos(\gamma)) J_0(\frac{\lambda}{M}\sqrt{l(l+1)}s) \nonumber \\
&&\simeq
\frac{1}{\pi^2}\sqrt{ \frac{M}{\lambda} } \frac{1}{s^{1/2}} \frac{1}{\sqrt{\sin\gamma}}\sum_{l=0}^{\infty}  \cos((l+\frac{1}{2})\gamma-\frac{\pi}{4})
\cos(\frac{\lambda}{M}(l+\frac{1}{2})s-\frac{\pi}{4}) \nonumber \\
&&=\frac{1}{2\pi^2}\sqrt{ \frac{M}{\lambda} } \frac{1}{s^{1/2}} \frac{1}{\sqrt{\sin\gamma}}\sum_{l=0}^{\infty} \left(  \cos((l+\frac{1}{2})(\gamma- \frac{\lambda}{M}s))+      \sin((l+\frac{1}{2})(\gamma+\frac{\lambda}{M}s))\right)
\, .
\lb{56}
\ee
Notice that we used the asymptotic forms for the Legendre and Bessel functions while keeping the sum over $l$ from $0$ to $\infty$.
This procedure is justified if we are interested in the singular part of the Green functions since the possible ``error'' is due to  to some non-singular  functions.

We then consider the sum over $l$ in (\ref{56}) as distribution and use the known formulas,
  \be
&& \sum_{l=0}^{\infty}  \cos((l+\frac{1}{2})x)=\pi \sum_{k=-\infty}^{+\infty} (-1)^k \delta(x+2\pi k)\, ,\nonumber \\
&&\sum_{l=0}^{\infty}  \sin((l+\frac{1}{2})x)=\sum_{k=-\infty}^{+\infty} (-1)^k  \operatorname{PV}\left({\frac{1}{x-2\pi k}}\right)
\, ,
\lb{57}
\ee
where $\operatorname{PV}(\frac{1}{x})$ denotes the Cauchy  principle value distribution. 
It is defined through its action on a test function $f(x)$ as
  \be
\left\langle \operatorname{PV}\frac{1}{x}, f \right\rangle
=
\lim_{\epsilon \to 0^+}
\left[
\int_{-\infty}^{-\epsilon} \frac{f(x)}{x}\,dx
+
\int_{\epsilon}^{\infty} \frac{f(x)}{x}\,dx
\right]
\, ,
\lb{58}
\ee
Our next observation is that the spacetime interval $s\geq 0$ and, hence, not all values of $k$ are allowed for singular terms in (\ref{56})
provided (\ref{57}) is used.
Taking this into account we finally arrive at the following form for the singular part of the Green function
  \be
&&G_R^{\rm sing}=\frac{1}{2\pi^2}\sqrt{ \frac{M}{\lambda} } \frac{1}{s^{1/2}} \frac{1}{\sqrt{\sin\gamma}}\, \Theta(t-t'-|z-z'|) \nonumber \\
&&\sum_{k\geq 0}\Big[ \pi(-1)^k \delta(\frac{\lambda}{M} s-\gamma -2\pi k)+(-1)^{k+1} \operatorname{PV}
\left(\frac{1}{\frac{\lambda}{M} s+\gamma -2\pi (k+1)} \right)\Big]\ \, .
\lb{60}
\ee
There are two sets of singularities in the Green function whose locations  are determined  by the conditions,
  \be
\frac{\lambda}{M}s=\gamma+2\pi k\,  \  \  {\rm and } \  \  \frac{\lambda}{M}s=-\gamma+2\pi (k+1)\, , \  k\geq 0
\lb{61}
\ee
where $s=\sqrt{(t-t')^2-(z-z')^2}$. These correspond to  the locations of the caustics in spacetime $M_{1,1}\times S^2$.
Equation (\ref{60}) reproduces the well-known four-fold singularity structure, discussed in earlier works,
giving rise to the sequence of singular terms 
$\delta(\sigma) \rightarrow \operatorname{PV} (\frac{1}{\sigma})\rightarrow -\delta(\sigma)\rightarrow -\operatorname{PV} (\frac{1}{\sigma})
$, where $\sigma^2=\lambda^2s^2-M^2(2\pi k\pm \gamma)^2$ is the four-dimensional spacetime interval.

\subsection{Caustics and the arrival times }\lb{arrivals}
The infinite series of singular terms in  (\ref{60}) arises due to the presence of caustics: the respective geodesics connecting two points on the two-sphere intersect at an infinite sequence of points.  In the sum in ((\ref{60}), the direct path between the two points on the sphere
corresponds to $k=0$, while the first crossing (caustic) occurs for $k=1$, and so on. As we mentioned above the type of singularity changes successively
for $k=0, 1, 2, 3$, after which  the same sequence is  repeated infinitely. 

Let us focus on the delta function singularities in (\ref{60}). These correspond to  perturbations propagating along   null geodesics that wrap several  times around the large circle of two-sphere $S^2$. Let us assume that the initial perturbation at $t=0$ is localized at a certain point on $S^2$  and is supported inside the interval $\left[-R,R\right]$. We place the observer at a large distance from the initial perturbation, $z\gg R$ and at the same point on the two-sphere.
The observer then sees a sequence of signals arriving at the observation point at a sequence of times $t_k$, which,   for sufficiently 
large  integer $k$, are given by equation
  \be
t_k=\sqrt{z^2+\frac{M^2}{\lambda^2}(2\pi k)^2}
\lb{62}
\ee
We note that  winding around the sphere introduces  a new, very large scale, $M/\lambda$, which is the time required for  light to traverse a the large circle of $S^2$. This is the largest scale   present in the wormhole geometry. The other large scale, $M\ln 1/\lambda$, is   much smaller for $\lambda \ll 1$. As a result, the arrival times $t_k$ parametrically very large.
This is why we  refer to this sequence as a {\it long-delay signal train}. It should be   distinguished  from the well-known sequence of echoes that appear in the non-extremal black hole mimickers, whose arrival times scale as $M\ln 1/\lambda$.

Two cases can be considered. First is when $z\gg 2\pi k M/\lambda$. In this case, one finds
  \be
t_k=z+\frac{M^2}{2\lambda^2 z}(2\pi k)^2
\lb{63}
\ee
Thus, the time interval between the two consecutive arrivals (recalling  that $z=\lambda \rho$) is
  \be
t_{k+1}-t_k=\frac{4\pi kM^2}{2\lambda \rho}\, , \ \ k\gg 1
\lb{64}
\ee
which grows with $k$.

The second case is when  $z\ll 2\pi k M/\lambda$. In this case, the arrival times are 
  \be
t_{k}=\frac{2\pi kM}{\lambda }\, , \ \ k\gg 1
\lb{65}
\ee
The interval between  two consecutive arrivals is then
  \be
t_{k+1}-t_k=\frac{2\pi M}{\lambda }\, , \ \ k\gg 1
\lb{66}
\ee
which is independent of $k$.  We see that the characteristic feature of the long-delay signal train is that
 the interval between consecutive signals is proportional to the inverse deformation parameter, $1/\lambda$.
An analysis that is rather similar can be done for the arrival times of the signal due to the PV singularities in the Green function
(\ref{60}).

\subsection{A  Gaussian perturbation in the throat}
Consider the case of the initial conditions $\Phi(t=0, z, \theta, \phi)=0$ and $\partial_t \Phi(t=0, z, \theta, \phi)=h(z, \theta, \phi)$.
The solution of the equation (\ref{23}) in the throat then takes the form
  \be
\Phi(t, z, \theta, \phi)=\int dz' d\phi' d \theta' \sin \theta' \, G_R(t,  z-z', \theta, \theta', \phi- \phi') h(z', \theta', \phi')
\lb{67}
\ee
In order to make our analysis more concrete and simplify the calculations consider the following choice for the initial function,
  \be
h(z', \theta', \phi')=e^{-\lambda z'^2} \frac{\delta(\theta')}{2\pi \sin \theta'}\, .
\lb{68}
\ee
It corresponds to a perturbation that is a delta-like spike at the north pole of the two-sphere and a Gaussian pulse in the $z$-direction concentrated at $z'=0$.  The respective geodesic distance defined by eq. (\ref{53}) between two points, $(\theta, \phi)$ and $(\theta', \phi')$,  on the two-sphere that appears in the integral (\ref{67}) is $\gamma=\theta$.
For the Green function we use the singular part (\ref{60}) skipping  thus all tail terms.

The calculation although tedious is straightforward. Here we present the final result.
  \be
&&\Phi(t, z, \theta, \phi)=\frac{1}{2\pi^2}\left(\frac{M}{\lambda}  \right)^{3/2}\frac{1}{\sqrt{\sin \theta}} 
\sum_{k\geq 0} \Big(\Theta(t-\tau_k(\theta))\, \pi\,  (-1)^k F_k(t,z,\theta)\nonumber \\
&&+\Theta(t-\tau_{k+1}(-\theta))\,   (-1)^{k+1} F_{k+1}(t,z,-\theta)\, 
\ln (\frac{t-\tau_{k+1}(-\theta)}{\tau_{k+1}(-\theta)})\Big) \, ,
\lb{69}
\ee
where $\tau_k(\theta)=\frac{M}{\lambda} (\theta+2\pi k)$ and we introduced function
  \be
F_k(t,z,\theta)=\sqrt{\frac{\tau_k(\theta)}{t^2-\tau^2_k(\theta)}}\left(  
e^{-\lambda(z+\sqrt{t^2-\tau_k^2(\theta)}\, )^2}+e^{-\lambda(z-\sqrt{t^2-\tau_k^2(\theta)}\, )^2}
       \right)\, .
\lb{70}
\ee
One technical remark is concerning the way we handle the  $\operatorname{PV}$ singularities in the above calculation.
The integral of interest is of the following form (here $0< s_k<t$)
\be
\operatorname{PV} \int_0^{t} ds \frac{f(s)}{s-s_k} =\operatorname{PV} \int_0^t ds \frac{f(s_k)}{s-s_k} + \int_0^t ds \frac{f(s)-f(s_k)}{s-s_k}\, .
\lb{70-1}
\ee
The second integral in the r.h.s. does not contain singularity and is not included in our result (\ref{70}). The singular integral is then computed as follows
\be
\operatorname{PV} \int_0^{t} ds \frac{1}{s-s_k} = \ln(\frac{t-s_k}{s_k})\, .
\lb{70-2}
\ee
This explains the origin of the 
 logarithmic terms in (\ref{69}).

The solution (\ref{69}) is a combination of two sequences, one is due to the delta-functions and the other is
due to the $\operatorname{PV}$-terms in the Green functions.
At a fixed $t$, only a finite number of integers $k$  appear in the sum (\ref{69}), namely those satisfying either  
$\tau_k(\theta)<t$ or $\tau_{k+1}(-\theta)<t$. 

At a fixed $z>0$ the solution (\ref{69}) represents a sequence of Gaussian signals whose maxima arrive at the times
$t_k(z,\theta)=\sqrt{z^2+\tau_k(\theta)}$ and $t'_k(z,\theta)=\sqrt{z^2+\tau_{k+1}(-\theta)}$. The general behavior of the arrival times is as we discussed in section \ref{arrivals}. The time between the two  successive  arrivals is very long, being  proportional to $M/\lambda$. 
The arrival time  depends on the angular coordinate $\theta$
on the two-sphere at the observation point. Consequently, we find a long delay in the arrival of the same Gaussian signal (with the integer $k$ fixed)
at two points one the two-sphere with  different azimuthal coordinates, while keeping the same value of coordinate $z$ at the observation point.
For example, the difference in the arrival times of the signal at the antipodal points $\theta=\pi$ and  $\theta=0$  satisfies equation
$t_k^2(\theta=\pi)-t_k^2(\theta=0)=4\pi^2M^2/\lambda^2(k+1/2)$.
Thus, the time delay  occurs not only along the radial direction ($z$), but also along the azimuthal direction on the two-sphere at fixed $z$.

Another notable feature  is that the Gaussian signals in (\ref{69})  flip sign for every second  signal in the delta- and PV-sequence.
This is the reflection of the four-fold  singularity structure of the Green function. 
Thus, the solution (\ref{69}) exhibits several peculiar features that could make it readily identifiable in a possible observational experiment.

\subsection{Signals detected by a distant observer}

A perturbation, once created in the throat, then propagates as a long-delay signal train toward large values of the coordinate $z$, where eventually the signals encounter the effective radial potential. The exact way in which they penetrate the barrier is an interesting problem that deserves to be studied. However, since the signal can be Fourier decomposed into modes $e^{-i\omega t}$, the part of the signal with sufficiently large $\omega$, above the maximum (\ref{31}) of the effective potential, will penetrate the barrier and potentially become detectable by a distant observer. The observer will see the long-delay signal train with the arrival times as discussed above. The characteristic time scale for the delay is $M/\lambda$.
This makes the signal train particularly distinguishable from the by now conventional echoes in black-hole mimickers, whose periodicity scales as $M \ln 1/\lambda$. It should be noted that for a mimicker of an extremal black hole, these conventional echoes are in fact absent. Indeed, they originate from the structure with two asymptotically flat regions and a symmetric two-barrier effective radial potential. In the present case of a mimicker of an extremal black hole, there is only one asymptotically flat region with only one maximum of the effective potential. Therefore, a signal sent to the mimicker from outside will not produce a sequence of periodic signals due to scattering off the maximum of the potential in a second asymptotic region.
Thus, in the case of an extremal black-hole mimicker, the detection of the long-delay signal train remains the only observable possibility to identify the mimicker as a no-horizon object. The features of the train that we described above (sign flips, azimuthal delays, etc.) are particularly instrumental in the observational identification of the mimicker.

\subsection{Generalization to wormhole mimickers of non-extremal black holes}

It is worth mentioning that the spacetime $M_{1,1}\times S^2$ also appears as an approximate geometry in the throat of a wormhole mimicker of a non-extremal black hole \cite{Damour:2007ap}. In this case, the approximation is less accurate than in the case of the mimicker of an extremal black hole. The metric $M_{1,1}\times S^2$ does not reproduce correctly the curvature in the throat since the deviation is quadratic in the radial coordinate $\rho$.
However, for a small deformation parameter, the throat is sufficiently long, and the approximation becomes relevant.
This suggests that a perturbation created in the throat of such a mimicker will undergo a time evolution similar to that described in the case considered here. Therefore, a long-delay signal train with the same features as those described above may also arise in the throat of a wormhole that mimics a non-extremal black hole, such as a Schwarzschild black hole. This will provide new observational possibilities for distinguishing such mimickers from black holes, in addition to the conventional echo effect.

\subsection{The AdS/CFT correspondence and unitarity }

In this section, we pause to make a comment that is not directly related to the main subject of this paper.

One of the motivations for introducing the wormhole deformation of the near-horizon region, as originally proposed in \cite{Solodukhin:2005qy}\footnote{It should be mentioned that \cite{Solodukhin:2005qy} 
uses the term ``Heisenberg time'' for what we denote here by $t_{\max}$. Our terminology here follows the more standard distinction between $t_{\max}$, the time at which finite-$N$ effects first become visible, and $t_H$, the much longer time scale associated with resolving the discreteness of the spectrum.}, was to reproduce the hierarchy of time scales characteristic of a system with large entropy. A large system in a finite volume with large entropy has a fundamentally discrete spectrum. However, if the system is observed only over a relatively short time, this discreteness is not apparent. It becomes resolvable only after a sufficiently long observation time. This time scale is the Heisenberg time, $t_H=1/<\delta \omega >$ where $<\delta \omega >$  is the average spacing between neighboring energy levels. Thus, for $t\sim t_H$, the discreteness of the spectrum can be resolved. For a system with large entropy, this is a parametrically large time scale, typically growing exponentially with the entropy, $t_H\sim e^S$.

A second characteristic time scale is what is commonly referred to in the literature as $t_{max}$. It characterizes the response of the system to a perturbation. For a system with a continuous spectrum, the response to a small perturbation is dissipative, as reflected in the complex poles of the corresponding two-point correlation function. The time $t_{max}$ is the time at which the system first becomes sensitive to the finiteness of its Hilbert space: the exponentially decaying semiclassical correlator becomes comparable to the finite-$N$ fluctuations.

A classical black hole, although it has finite entropy $S$, exhibits only this dissipative behavior. In particular, there is no corresponding late-time restoration of the initial perturbation. If the complete evolution is unitary, however, one expects that a black hole initially disturbed by a perturbation should eventually exhibit a restoration of information about that perturbation, at sufficiently long times of order $t_H\sim e^S$. The absence of such behavior in the classical black-hole description is one manifestation of the information-loss problem.

It is therefore interesting to ask whether this puzzle can be understood within the AdS/CFT correspondence, which is valid in an appropriate large-$N$ limit, provided that unitarity is restored at finite $N$. Notice that, owing to the AdS boundary conditions, the dual system effectively lives in a finite volume. The earlier works in this direction include \cite{Dyson:2002pf}-\cite{Bairy:2026zsk}.

Geometrically, the origin of the dissipative behavior of black holes is the presence of the horizon. Near the horizon, perturbations effectively propagate through an infinite volume of the optical metric. This feature is common to black holes, including those in asymptotically AdS spacetimes. Replacing the horizon by a wormhole throat makes this effective optical volume finite, although parametrically large. Such a modification necessarily involves a non-perturbative deformation of the metric.

Within this picture, provided the deformation parameter $\lambda\sim e^S$, the time scale $t_{max}$ can be identified with the time required for a light signal to probe the throat. The size of the throat, and hence the corresponding time scale, grows logarithmically with the deformation parameter, $t_{max}\sim \ln (1/\lambda)$.
The relevant modes at this stage are those propagating radially along the throat. In the throat region, the approximate geometry is
 $M_{1,1}\times S^{d-2}$ 
for a $d$-dimensional spacetime. The time coordinate in the throat is related to the asymptotic time by $t_{th}=\lambda t$.

The present study suggests that the modes relevant for the much later recovery of information about the initial perturbation are instead modes that propagate along the periodic geodesics of the $(d-2)$-sphere, in addition to their propagation along the throat in the radial direction. As discussed in detail above, the time required for such a geodesic to complete a circuit of the sphere is parametrically of 
order $1/\lambda$. We therefore identify this scale as a geometrical candidate for the Heisenberg time, $t_H\sim 1/\lambda$.
This gives the desired hierarchy of time scales, $t_{max}\sim \ln (1/\lambda)$ and $t_H\sim 1/\lambda$.
The first scale corresponds to the time required to probe the finite throat, whereas the second arises from the much longer periodic propagation in the angular directions.

A special case is the $d=3$ BTZ geometry. The wormhole modification of a BTZ black hole was introduced in \cite{Solodukhin:2005qy}. In this case, the ``spherical'' part is a circle $S^1$. There are no caustics in this case.  
Nevertheless, the long timescale $1/\lambda$ associated with periodic propagation is already present. The four-fold singularity structure is therefore a separate effect, associated with caustic formation on $S^{d-2}$ for $d>3$, and should not be conflated with the origin of the $1/\lambda$ timescale itself.

For the purposes of the argument in this section, the deformation parameter is assumed to scale as $\lambda\sim e^{-S}$. However, throughout the rest of the paper, $\lambda$ is treated as a free deformation parameter to be constrained by observational data.

\section{Mimicker of an extremal Kerr black hole}
\setcounter{equation}0
In this section we extend the construction of extremal black hole mimicker to rotating case. 
A metric that describes a wormhole modification of the Kerr black hole, characterized by mass $M$ and rotation $a$, was suggested in \cite{Bueno:2017hyj}. Their prescription is to keep same mass parameter in all components of the metric except for $(r,r)$
component as in the Kerr metric and only in $(r,r)$ component replace the mass parameter with $M_1>M$. In the extremal case mass is related to rotation, $M=a$. So that there remains only one parameter in the extremal Kerr metric. Our prescription for the extremal case
is then take the rotating wormhole metric of  \cite{Bueno:2017hyj}, that effectively has three parameters $M$, $a$ and $M_2=M(1+\lambda)$,
and take the extremal limit as $M=a$. We then obtain the following metric (for esthetic reasons we use $a=M$ and $b=M_2$, $b>a$),
  \be
ds^2=-(1-\frac{2a r}{\Sigma} )dt^2
-\frac{4a^2 r\sin^2\theta}{\Sigma} dt d\phi +\frac{\Sigma}{\Delta (r)}dr^2+\Sigma d\theta^2+(r^2+a^2+\frac{2a^3 r}{\Sigma})\sin^2 \theta d\phi^2\, ,
\lb{71}
\ee
where
\be
\Sigma=r^2+a^2\cos^2\theta\, , \  \  \Delta(r)=(r-b)^2\, .
\lb{72}
\ee
The throat is located at $r=b$.  The small deformation parameter now is 
\be
\lambda=\frac{b-a}{2a}>0\, .
\lb{72-X}
\ee
The scalar curvature takes a finite value at the throat,
\be
R=\frac{2(\cos^4\theta a^4+a^2(a^2-2ab+4b^2)\cos^2\theta +b^2(b^2-a^2))}{(b^2+a^2\cos^2\theta)^3}\, .
\lb{73}
\ee
As in the static case, in the throat,  it is convenient introduce a radial coordinate $\rho=b\ln(r-b)$. The throat then corresponds to large negative values of $\rho$. 
The metric at the throat then takes the form
  \be
ds^2_{\rm th}=\frac{dl_1dl_2}{(b^2+a^2\cos^2\theta )}+(b^2+a^2\cos^2\theta)(b^{-2}d\rho^2+d\theta^2)
\lb{74}
\ee
We introduced $dl_1$ and $dl_2$  by relations 
\be
&&dl_1+dl_2=2(a^2+b^2)\sin\theta d\hat{\phi}\, , \nonumber  \\  
&&dl_1-dl_2=4(\frac{b^2+a^2\cos^2\theta}{b^2+a^2} )ad\hat{t}-2(b-a)a\sin^2\theta d\hat{\phi}\, ,
\lb{75}
\ee
where we introduced 
\be
\hat{t}=(\frac{b-a}{2a})t\, , \ \  \hat{\phi}=\phi-\frac{a}{a^2+b^2}t
\lb{76}
\ee
Metric (\ref{74}) has a smooth $b=a$ limit (keeping $b-a$ finite in $\hat{t}$),
\be
ds^2_{\rm th}=(1+\cos^2\theta)(-d\hat{t}^2+d\rho^2)+a^2\left((1+\cos^2\theta) d\theta^2+\frac{4\sin^2\theta d\hat{\phi}^2}{1+\cos^2\theta}\right)\, .
\lb{77}
\ee
It should be noted a certain similarity of the throat metric (\ref{77}) and the near-horizon limit of the extremal black hole studied in \cite{Bardeen:1999px}.
The important difference is the appearance of $AdS_2$ in the near-horizon limit while in the throat of the mimicker there appears Minkowski  spacetime $M_{1,1}$.

The scalar curvature of metric (\ref{77}) 
\be
R=\frac{2\cos^2\theta(\cos^2+3)}{a^2(1+\cos^2\theta)^3}
\lb{78}
\ee
that is precisely the $a=b$ limit of curvature (\ref{74}). The metric (\ref{77}) is a warped product of two dimensional Minkowski spacetime 
$M_{1,1}$ with coordinates $(\hat{t}, \rho)$ and a compact two-surface $\Sigma_2$. This spacetime replaces the product spacetime
$M_{1,1}\times S^2$ in the static case. Notice that the throat time $\hat{t}$ is related to the time $t$ at infinity as $\hat{t}=\lambda t$
where $\lambda=(b-a)/2a$ is the deformation parameter. It is similar to what we had in the throat in static metric (\ref{14}).

Since for null geodesics the spacetime interval vanishes their analysis can be done in a conformally related metric
\be
ds^2_{\rm opt}=-d\hat{t}^2+d\rho^2+a^2\left(d\theta^2+\frac{4\sin^2\theta d\hat{\phi}^2}{(1+\cos^2\theta)^2}\right)\, .
\lb{79}
\ee
The two-dimensional part of the metric $(\theta,\hat{\phi})$ has scalar curvature 
\be
R_\Sigma=\frac{2(\cos(\theta)^4 - 12\cos(\theta)^2 + 3)}{a^2(1 + \cos(\theta)^2)^2}
\lb{80}
\ee
that changes sign from negative value $-4/a^2$ at the poles $\theta=0$ and $ \theta=\pi$ to positive value $6/a^2$ at the equator
$\theta=\pi/2$.

Let us analyze whether  in the metric (\ref{79}) there exist circular null orbits or more generally the closed orbits of null geodesics.
We fix the radial coordinate $\rho=const$ then the null geodesic $(\hat{t}(\sigma), \theta(\sigma), \phi(\sigma))$
is described by the equations
\be
\dot{\hat{t}}=E\, , \  \    \dot{\hat{\phi}}=\frac{L}{a^2f(\theta)} \, , \  \   (\dot{{\theta}})^2=\frac{E^2}{a^2}\Big(1-\frac{q^2}{f(\theta)}\Big)
\, , \  \   f(\theta)=\frac{4\sin^2\theta }{(1+\cos^2\theta)^2}
\lb{81}
\ee
where we introduced parameter $q=L/(aE)$. In order to have a solution to equation for the azimuthal angle one get the condition: 
$0\leq q \leq 2$.

{For $q=0$} ($L=0$), the geodesic is meridional and follows a curve of constant azimuthal angle $\hat{\phi}={\rm const}$.  
A family of such geodesics emitted
 from the north pole ($\theta=0$) with different values of $\hat{\phi}$  all converge at the south pole ($\theta=\pi$),  forming a caustic. 
The time required to travel from the north pole to the south pole is $\Delta\hat{t}_{NS}=\pi a$ independently of $E$ and of the  initial value of $\hat{\phi}$.
Upon continuation through the poles, the meridional geodesic becomes a closed, great-circle-like spatial geodesic. 
 The corresponding travel  is
$\Delta\hat{t}_{NN}=2\pi a$ or, in terms of the time   $t$ measured 	at infinity, $\Delta t_{NN}=2\pi a/\lambda$. This time becomes very 
large for 
small deformation parameter $\lambda$. On the other hand, the co-rotating azimuthal coordinate $\hat{\phi}$ is related to the azimuthal coordinate $\phi$
by (\ref{76}). Therefore, for $b$ close to $a$ one finds that after one complete meridional cycle  the azimuthal angle changes by
$\Delta \phi_{NN}=\Delta t_{NN}/(2a)=\pi/\lambda$.  Since $\lambda\ll 1$, the photon therefore makes  a large number of 
azimuthal revolutions around the Kerr symmetry axis before returning to the same polar position.
The trajectory closes in the original azimuthal coordinate 
 if $1/(2\lambda)$ is rational; in particular it closes if it is an integer.  If $1/(2\lambda)$ is irrational the trajectory is quasi-periodic in the azimuthal direction. This provides  a simple mechanism for  long-delay behavior: one short closed orbit in the near-horizon optical geometry becomes a highly wound orbit when mapped back to the asymptotic Kerr azimuth, with a return time scaling like $1/\lambda$.

Note that our throat metric (\ref{77}), as well as its optical version (\ref{79}), is invariant under translations in the radial direction $\rho$. Consequently, the structure of the meridional null geodesics exists at every value of $\rho$. Moreover, one can add a nonzero radial velocity, so that the null geodesic becomes a superposition of the rotational motion on the two-surface and uniform propagation in the radial direction. This is analogous to the situation in the static case.
These geodesics can produce, for a field perturbation propagating through the throat, a sequence of long-delay signals with a delay time of order $\Delta t\sim 1/\lambda$, 
 similarly to the static case. The important difference, however, is that in the present case the sequence of signals is associated with propagation along the North-South meridional direction and therefore has a characteristic polarization aligned with the North-South direction.

\medskip

{ For value $q=2$} the trajectory is equatorial circle ($\theta=\pi/2$). One complete revolution takes $\Delta \hat{t}=E\sigma$, the change in azimuthal direction is
$\Delta\hat{\phi}=E\sigma/2a$. Provided the azimuthal coordinate $\Delta\phi=2\pi$ one finds that in terms of time of a distant 
observer  it takes   $\Delta t=4\pi a/(1+\lambda)\simeq 4\pi a$ for one complete revolution. The equatorial revolution is, thus,  much faster than the meridional one.

\medskip

For  $0<q< 2$ we have an interesting family of trajectories that oscillate between the two values $\theta_m$ and $\pi-\theta_m$
of the polar angle, where $\theta_m$ is determined by the turning-point condition
\be
\cos^2\theta_m=\frac{2\sqrt{2q^2+1}-q^2-2}{q^2} \, .
\lb{82}
\ee
During one complete oscillation, the photon follows the trajectory $\theta_m\rightarrow \pi/2 \rightarrow  \pi-\theta_m \rightarrow \pi/2 \rightarrow \theta_m$.  The corresponding change in the azimuthal angle is
\be
\Delta \hat{\phi}=4q\int_{\theta_m}^{\pi/2} \frac{d\theta}{\sqrt{f(\theta)(f(\theta)-q^2)}}=\Upsilon (q)
\lb{83}
\ee
while the corresponding propagation time is
\be
\Delta \hat{t}=4a\int_{\theta_m}^{\pi/2} \frac{\sqrt{f(\theta)}d\theta}{\sqrt{f(\theta)-q^2}}= 2a\mathrm{T} (q)
\lb{84}
\ee
Thus, $\Upsilon (q)$  gives the azimuthal advance, while $\mathrm{T} (q)$ gives the time required for one complete polar oscillation.
In terms of time $t=\hat{t}/\lambda$ measured at infinity one finds $\Delta t=2a\mathrm{T} (q)/\lambda$ for one such oscillation. This time is
very large for $\lambda \ll 1$.
Using the relation between the co-rotating and Kerr azimuthal coordinates given in (\ref{76}), the corresponding change in the Kerr azimuthal angle $\phi=\hat{\phi}+t/2a$ is
\be
\Delta\phi= \Upsilon (q)+\frac{1}{\lambda} \mathrm{T} (q)\, .
\lb{85}
\ee
Provided $\Delta\phi/2\pi= l/p$ is rational the respective trajectory is periodic. In this case, the trajectory closes after $p$ polar oscillations and $l$ complete azimuthal revolutions. If  $\Delta\phi/2\pi$ is irrational, which appears to be the more generic situation given the elliptic-type integrals encountered above, the trajectory is instead quasi-periodic.

\section{Conclusions}
\setcounter{equation}0

We introduced a horizonless deformation (mimicker) of the extremal black hole and analyzed its geometry and scalar-field dynamics. The mimicker spacetime consists of an infinite throat at one end and an asymptotically flat region at the other. The geometry in the throat is closely described by the spacetime $M_{1,1}\times S^2$. It has several interesting properties, which we have discussed throughout this paper. Among them is the existence of a family of circular null geodesics. The corresponding characteristic time scale is of order $1/\lambda$, where $\lambda$ is the deformation parameter. This provides a new time scale that has not previously been discussed in the context of black-hole mimickers.

A much shorter time scale, of order $\ln(1/\lambda)$, is associated with the double-peak structure of the radial potential in mimickers of non-extremal black holes and sets the characteristic time scale for the echo effect. In the case of the extremal black-hole mimicker, the radial potential, as in the extremal black hole itself, has only one peak, and the conventional echo effect is absent.

The caustics produced by the periodic null geodesics in the throat give rise to an infinite sequence of singular terms in the Green function for a scalar field. As a result, a scalar-field perturbation in the throat produces a sequence of long-delay signals, with a time interval of order  $1/\lambda$ between consecutive signals. The caustics generate the well-known fourfold structure of the singularities of the retarded Green function, leading to distinctive features in the signals, such as a sign flip in every second signal in the sequence.

We extended this picture to the rotating case. For a mimicker of the extremal Kerr black hole, we analyzed the corresponding throat geometry and found a family of null geodesics that produce caustics, similarly to the static case. This leads to a sequence of long-delay signals, with additional features arising from rotation, which we discussed.

We also investigated the stability of scalar-field dynamics in the mimicker spacetime. We found a spatial-derivative instability that resembles the Aretakis instability but has a different origin, being associated with the non-smoothness of the initial data rather than with the horizon geometry. On the other hand, a local Aretakis-like analysis near the throat produces apparently exponentially growing solutions. We show, however, that these do not correspond to physical unstable modes of the full spacetime. The global radial problem has a non-negative effective potential, and the appropriate boundary conditions exclude exponentially growing modes. 
An interesting question for future work is whether the mimicker considered here could represent an endpoint of the Aretakis instability.

We expect that many of the aspects and phenomena uncovered in the present paper should also persist, to some extent, for mimickers of non-extremal black holes. In this case, the spacetime $M_{1,1}\times S^2$ provides an approximate description of the geometry within the long but finite throat. This may lead to additional potentially observable manifestations of the horizonless nature of mimickers, as well as to the emergence of the new time scale $1/\lambda$. These and other related issues warrant further investigation.

\bigskip
\section*{Acknowledgements}  I would like to thank D. Sarkar, V. Tagiev, M. Volkov  and O. Zaslavski for helpful remarks.


\newpage
\begin{thebibliography}{999}
{\frenchspacing \parskip=2mm

\bibitem{BHreview}
R.~ Genzel, F.~ Eisenhauer and S.~ Gillessen,  ``Experimental studies of black holes: status and future prospects,''  Astron. Astrophys.
 Rev. \textbf{32}, 3 (2024). 

\bibitem{Damour:2007ap}
T.~Damour and S.~N.~Solodukhin,
``Wormholes as black hole foils,''
Phys. Rev. D \textbf{76} (2007), 024016
\href{https://arxiv.org/abs/0704.2667}{[arXiv:0704.2667 [gr-qc]]}.

\bibitem{Solodukhin:2005qy}
S.~N.~Solodukhin,
``Restoring unitarity in BTZ black hole,''
Phys. Rev. D \textbf{71} (2005), 064006
[arXiv:hep-th/0501053 [hep-th]].

\bibitem{Bueno:2017hyj}
P.~Bueno, P.~A.~Cano, F.~Goelen, T.~Hertog and B.~Vercnocke,
``Echoes of Kerr-like wormholes,''
Phys. Rev. D \textbf{97} (2018) no.2, 024040
\href{https://arxiv.org/abs/1711.00391}{[arXiv:1711.00391 [gr-qc]]}.


\bibitem{Aretakis:2011ha}
S.~Aretakis,
``Stability and Instability of Extreme Reissner-Nordstr{\"o}m Black Hole Spacetimes for Linear Scalar Perturbations I,''
Commun. Math. Phys. \textbf{307} (2011), 17-63
[arXiv:1110.2007 [gr-qc]].

\bibitem{Aretakis:2011hc}
S.~Aretakis,
``Stability and Instability of Extreme Reissner-Nordstr{\"o}m Black Hole Spacetimes for Linear Scalar Perturbations II,''
Annales Henri Poincare \textbf{12} (2011), 1491-1538
[arXiv:1110.2009 [gr-qc]].

\bibitem{Lucietti:2012xr}
J.~Lucietti, K.~Murata, H.~S.~Reall and N.~Tanahashi,
``On the horizon instability of an extreme Reissner-Nordstr{\"o}m black hole,''
JHEP \textbf{03} (2013), 035
[arXiv:1212.2557 [gr-qc]].

\bibitem{Lucietti:2012sf}
J.~Lucietti and H.~S.~Reall,
``Gravitational instability of an extreme Kerr black hole,''
Phys. Rev. D \textbf{86} (2012), 104030
[arXiv:1208.1437 [gr-qc]].


\bibitem{Agrawal:2026oka}
S.~Agrawal, P.~Charalambous, L.~Donnay, S.~Liberati and G.~Neri,
``Taming the Aretakis instability: extremal black holes with multi-degenerate horizons,''
[arXiv:2604.15260 [gr-qc]].


\bibitem{Katagiri:2021scx}
T.~Katagiri and M.~Kimura,
``Revisiting the Aretakis constants and instability in two-dimensional anti{\textendash}de Sitter spacetimes,''
Phys. Rev. D \textbf{103} (2021) no.6, 064011
[arXiv:2101.03479 [gr-qc]].



\bibitem{Cardoso:2016rao}
V.~Cardoso, E.~Franzin and P.~Pani,
``Is the gravitational-wave ringdown a probe of the event horizon?,''
Phys. Rev. Lett. \textbf{116} (2016) no.17, 171101
[erratum: Phys. Rev. Lett. \textbf{117} (2016) no.8, 089902]
[arXiv:1602.07309 [gr-qc]].

\bibitem{Solodukhin:2025opw}
S.~N.~Solodukhin and V.~Tagiev,
``Exploring the landscape of black hole mimickers,''
Phys. Rev. D \textbf{113} (2026) no.6, 064017
[arXiv:2511.03879 [gr-qc]].

\bibitem{Solodukhin:2026vtz}
S.~N.~Solodukhin and V.~Tagiev,
``Tidal Love numbers of wormholes as black-hole mimickers,''
[arXiv:2609.13129 [gr-qc]].


\bibitem{Casals:2012px}
M.~Casals and B.~C.~Nolan,
``A Kirchhoff integral approach to the calculation of Green's functions beyond the normal neighbourhood,''
Phys. Rev. D \textbf{86} (2012), 024038
[arXiv:1204.0407 [gr-qc]].

\bibitem{Aruquipa:2022rwr}
D.~Q.~Aruquipa and M.~Casals,
``Hadamard tail from initial data on the light cone,''
Phys. Rev. D \textbf{107} (2023) no.8, 084008
[arXiv:2205.13677 [gr-qc]].


\bibitem{Reynolds:2020jwt}
C.~S.~Reynolds,
``Observational Constraints on Black Hole Spin,''
Ann. Rev. Astron. Astrophys. \textbf{59} (2021), 117-154
[arXiv:2011.08948 [astro-ph.HE]].



\bibitem{Berthiere:2017tms}
C.~Berthiere, D.~Sarkar and S.~N.~Solodukhin,
Phys. Lett. B \textbf{786} (2018), 21-27
\href{https://arxiv.org/abs/1712.09914}{[arXiv:1712.09914 [hep-th]]}.

\bibitem{Potaux:2021yan}
Y.~Potaux, D.~Sarkar and S.~N.~Solodukhin,
Phys. Rev. D \textbf{105} (2022) no.2, 025015
\href{https://arxiv.org/abs/2112.03855}{[arXiv:2112.03855 [hep-th]]}.

\bibitem{Potaux:2022uxa}
Y.~Potaux, D.~Sarkar and S.~N.~Solodukhin,
Phys. Rev. Lett. \textbf{130} (2023) no.26, 261501
\href{https://arxiv.org/abs/2212.13208}{[arXiv:2212.13208 [hep-th]]}.

\bibitem{Potaux:2023fwm}
Y.~Potaux, D.~Sarkar and S.~N.~Solodukhin,
Phys. Rev. D \textbf{108} (2023) no.12, 125012
\href{https://arxiv.org/abs/2310.18745}{[arXiv:2310.18745 [hep-th]]}.

\bibitem{Volkov:1998cc}
M.~S.~Volkov and D.~V.~Gal'tsov,
Phys. Rept. \textbf{319} (1999), 1-83
[arXiv:hep-th/9810070 [hep-th]].

\bibitem{Lemos:2008cv}
J.~P.~S.~Lemos and O.~B.~Zaslavskii,
Phys. Rev. D \textbf{78} (2008), 024040
[arXiv:0806.0845 [gr-qc]].

\bibitem{Sussman} E.~Sussman, ``Massive Wave Propagation Near Null Infinity, ''  Ann. PDE {\bf 12}, 9 (2026).

\bibitem{Winicour:1988aq}
J.~Winicour,
``MASSIVE FIELDS AT NULL INFINITY,''
J. Math. Phys. \textbf{29} (1988), 2117-212

\bibitem{Dyson:2002pf}
L.~Dyson, M.~Kleban and L.~Susskind,
``Disturbing implications of a cosmological constant,''
JHEP \textbf{10} (2002), 011
[arXiv:hep-th/0208013 [hep-th]].

\bibitem{Dyson:2002nt}
L.~Dyson, J.~Lindesay and L.~Susskind,
``Is there really a de Sitter/CFT duality?,''
JHEP \textbf{08} (2002), 045
[arXiv:hep-th/0202163 [hep-th]].

\bibitem{Barbon:2003aq}
J.~L.~F.~Barbon and E.~Rabinovici,
JHEP \textbf{11} (2003), 047
doi:10.1088/1126-6708/2003/11/047
[arXiv:hep-th/0308063 [hep-th]].

\bibitem{Solodukhin:2004rv}
S.~N.~Solodukhin,
``Can black hole relax unitarily?,''
[arXiv:hep-th/0406130 [hep-th]]. 

\bibitem{Kleban:2004rx}
M.~Kleban, M.~Porrati and R.~Rabadan,
``Poincare recurrences and topological diversity,''
JHEP \textbf{10} (2004), 030
[arXiv:hep-th/0407192 [hep-th]].

\bibitem{Bairy:2026zsk}
R.~Bairy, M.~Nath and D.~Sarkar,
``Near-horizon modifications in finite $N$ holography,''
[arXiv:2606.21537 [hep-th]].

\bibitem{Bardeen:1999px}
J.~M.~Bardeen and G.~T.~Horowitz,
``The Extreme Kerr throat geometry: A Vacuum analog of AdS(2) x S**2,''
Phys. Rev. D \textbf{60} (1999), 104030
[arXiv:hep-th/9905099 [hep-th]].




}



\end{thebibliography}
\end{document}